\documentclass[aps,prc,superscriptaddress,showpacs,nofootinbib,floatfix,twocolumn]{revtex4}

\usepackage{graphicx}	% Include figure files
\usepackage{subfigure}

\newcommand{\pt}{{p_{\rm T}}}
\newcommand{\dau}{d+{\rm Au}}

\newcommand{\auau}{{\rm Au+Au}}
\newcommand{\cucu}{{\rm Cu+Cu}}
\newcommand{\pp}{p+p}
\newcommand{\jpsi}{J/\psi}
\newcommand{\ncol}{{N_{\rm coll}}}
\newcommand{\rda}{{R_d{\rm Au}}}

\begin{document}

\title{Cold Nuclear Matter Effects on $J/\psi$ Production as Constrained 
by Deuteron-Gold Measurements at $\sqrt{s_{NN}}=200$~GeV}

\newcommand{\abilene}{Abilene Christian University, Abilene, TX 79699, U.S.}
\newcommand{\acadsin}{Institute of Physics, Academia Sinica, Taipei 11529, Taiwan}
\newcommand{\banaras}{Department of Physics, Banaras Hindu University, Varanasi 221005, India}
\newcommand{\barc}{Bhabha Atomic Research Centre, Bombay 400 085, India}
\newcommand{\bnl}{Brookhaven National Laboratory, Upton, NY 11973-5000, U.S.}
\newcommand{\caucr}{University of California - Riverside, Riverside, CA 92521, U.S.}
\newcommand{\charlesczech}{Charles University, Ovocn\'{y} trh 5, Praha 1, 116 36, Prague, Czech Republic}
\newcommand{\ciae}{China Institute of Atomic Energy (CIAE), Beijing, People's Republic of China}
\newcommand{\cns}{Center for Nuclear Study, Graduate School of Science, University of Tokyo, 7-3-1 Hongo, Bunkyo, Tokyo 113-0033, Japan}
\newcommand{\colorado}{University of Colorado, Boulder, CO 80309, U.S.}
\newcommand{\columbia}{Columbia University, New York, NY 10027 and Nevis Laboratories, Irvington, NY 10533, U.S.}
\newcommand{\czechtech}{Czech Technical University, Zikova 4, 166 36 Prague 6, Czech Republic}
\newcommand{\dapnia}{Dapnia, CEA Saclay, F-91191, Gif-sur-Yvette, France}
\newcommand{\debrecen}{Debrecen University, H-4010 Debrecen, Egyetem t{\'e}r 1, Hungary}
\newcommand{\elte}{ELTE, E{\"o}tv{\"o}s Lor{\'a}nd University, H - 1117 Budapest, P{\'a}zm{\'a}ny P. s. 1/A, Hungary}
\newcommand{\fit}{Florida Institute of Technology, Melbourne, FL 32901, U.S.}
\newcommand{\fsu}{Florida State University, Tallahassee, FL 32306, U.S.}
\newcommand{\gsu}{Georgia State University, Atlanta, GA 30303, U.S.}
\newcommand{\hiroshima}{Hiroshima University, Kagamiyama, Higashi-Hiroshima 739-8526, Japan}
\newcommand{\ihepprot}{IHEP Protvino, State Research Center of Russian Federation, Institute for High Energy Physics, Protvino, 142281, Russia}
\newcommand{\illuiuc}{University of Illinois at Urbana-Champaign, Urbana, IL 61801, U.S.}
\newcommand{\instpasczech}{Institute of Physics, Academy of Sciences of the Czech Republic, Na Slovance 2, 182 21 Prague 8, Czech Republic}
\newcommand{\isu}{Iowa State University, Ames, IA 50011, U.S.}
\newcommand{\jinrdubna}{Joint Institute for Nuclear Research, 141980 Dubna, Moscow Region, Russia}
\newcommand{\kek}{KEK, High Energy Accelerator Research Organization, Tsukuba, Ibaraki 305-0801, Japan}
\newcommand{\kfki}{KFKI Research Institute for Particle and Nuclear Physics of the Hungarian Academy of Sciences (MTA KFKI RMKI), H-1525 Budapest 114, POBox 49, Budapest, Hungary}
\newcommand{\korea}{Korea University, Seoul, 136-701, Korea}
\newcommand{\kurchatov}{Russian Research Center ``Kurchatov Institute", Moscow, Russia}
\newcommand{\kyoto}{Kyoto University, Kyoto 606-8502, Japan}
\newcommand{\labllr}{Laboratoire Leprince-Ringuet, Ecole Polytechnique, CNRS-IN2P3, Route de Saclay, F-91128, Palaiseau, France}
\newcommand{\lawllnl}{Lawrence Livermore National Laboratory, Livermore, CA 94550, U.S.}
\newcommand{\losalamos}{Los Alamos National Laboratory, Los Alamos, NM 87545, U.S.}
\newcommand{\lpc}{LPC, Universit{\'e} Blaise Pascal, CNRS-IN2P3, Clermont-Fd, 63177 Aubiere Cedex, France}
\newcommand{\lund}{Department of Physics, Lund University, Box 118, SE-221 00 Lund, Sweden}
\newcommand{\muenster}{Institut f\"ur Kernphysik, University of Muenster, D-48149 Muenster, Germany}
\newcommand{\myongji}{Myongji University, Yongin, Kyonggido 449-728, Korea}
\newcommand{\nagasaki}{Nagasaki Institute of Applied Science, Nagasaki-shi, Nagasaki 851-0193, Japan}
\newcommand{\newmex}{University of New Mexico, Albuquerque, NM 87131, U.S. }
\newcommand{\nmsu}{New Mexico State University, Las Cruces, NM 88003, U.S.}
\newcommand{\ornl}{Oak Ridge National Laboratory, Oak Ridge, TN 37831, U.S.}
\newcommand{\orsay}{IPN-Orsay, Universite Paris Sud, CNRS-IN2P3, BP1, F-91406, Orsay, France}
\newcommand{\peking}{Peking University, Beijing, People's Republic of China}
\newcommand{\pnpi}{PNPI, Petersburg Nuclear Physics Institute, Gatchina, Leningrad region, 188300, Russia}
\newcommand{\riken}{RIKEN, The Institute of Physical and Chemical Research, Wako, Saitama 351-0198, Japan}
\newcommand{\rikjrbrc}{RIKEN BNL Research Center, Brookhaven National Laboratory, Upton, NY 11973-5000, U.S.}
\newcommand{\rikkyo}{Physics Department, Rikkyo University, 3-34-1 Nishi-Ikebukuro, Toshima, Tokyo 171-8501, Japan}
\newcommand{\saispbstu}{Saint Petersburg State Polytechnic University, St. Petersburg, Russia}
\newcommand{\saopaulo}{Universidade de S{\~a}o Paulo, Instituto de F\'{\i}sica, Caixa Postal 66318, S{\~a}o Paulo CEP05315-970, Brazil}
\newcommand{\seoulnat}{System Electronics Laboratory, Seoul National University, Seoul, Korea}
\newcommand{\stonybrkc}{Chemistry Department, Stony Brook University, Stony Brook, SUNY, NY 11794-3400, U.S.}
\newcommand{\stonycrkp}{Department of Physics and Astronomy, Stony Brook University, SUNY, Stony Brook, NY 11794, U.S.}
\newcommand{\subatech}{SUBATECH (Ecole des Mines de Nantes, CNRS-IN2P3, Universit{\'e} de Nantes) BP 20722 - 44307, Nantes, France}
\newcommand{\tenn}{University of Tennessee, Knoxville, TN 37996, U.S.}
\newcommand{\titech}{Department of Physics, Tokyo Institute of Technology, Oh-okayama, Meguro, Tokyo 152-8551, Japan}
\newcommand{\tsukuba}{Institute of Physics, University of Tsukuba, Tsukuba, Ibaraki 305, Japan}
\newcommand{\vandy}{Vanderbilt University, Nashville, TN 37235, U.S.}
\newcommand{\waseda}{Waseda University, Advanced Research Institute for Science and Engineering, 17 Kikui-cho, Shinjuku-ku, Tokyo 162-0044, Japan}
\newcommand{\weizmann}{Weizmann Institute, Rehovot 76100, Israel}
\newcommand{\yonsei}{Yonsei University, IPAP, Seoul 120-749, Korea}
\affiliation{\abilene}
\affiliation{\acadsin}
\affiliation{\banaras}
\affiliation{\barc}
\affiliation{\bnl}
\affiliation{\caucr}
\affiliation{\charlesczech}
\affiliation{\ciae}
\affiliation{\cns}
\affiliation{\colorado}
\affiliation{\columbia}
\affiliation{\czechtech}
\affiliation{\dapnia}
\affiliation{\debrecen}
\affiliation{\elte}
\affiliation{\fit}
\affiliation{\fsu}
\affiliation{\gsu}
\affiliation{\hiroshima}
\affiliation{\ihepprot}
\affiliation{\illuiuc}
\affiliation{\instpasczech}
\affiliation{\isu}
\affiliation{\jinrdubna}
\affiliation{\kek}
\affiliation{\kfki}
\affiliation{\korea}
\affiliation{\kurchatov}
\affiliation{\kyoto}
\affiliation{\labllr}
\affiliation{\lawllnl}
\affiliation{\losalamos}
\affiliation{\lpc}
\affiliation{\lund}
\affiliation{\muenster}
\affiliation{\myongji}
\affiliation{\nagasaki}
\affiliation{\newmex}
\affiliation{\nmsu}
\affiliation{\ornl}
\affiliation{\orsay}
\affiliation{\peking}
\affiliation{\pnpi}
\affiliation{\riken}
\affiliation{\rikjrbrc}
\affiliation{\rikkyo}
\affiliation{\saispbstu}
\affiliation{\saopaulo}
\affiliation{\seoulnat}
\affiliation{\stonybrkc}
\affiliation{\stonycrkp}
\affiliation{\subatech}
\affiliation{\tenn}
\affiliation{\titech}
\affiliation{\tsukuba}
\affiliation{\vandy}
\affiliation{\waseda}
\affiliation{\weizmann}
\affiliation{\yonsei}
\author{A.~Adare}	\affiliation{\colorado}
\author{S.S.~Adler}	\affiliation{\bnl}
\author{S.~Afanasiev}	\affiliation{\jinrdubna}
\author{C.~Aidala}	\affiliation{\columbia}
\author{N.N.~Ajitanand}	\affiliation{\stonybrkc}
\author{Y.~Akiba}	\affiliation{\kek}  \affiliation{\riken}  \affiliation{\rikjrbrc}
\author{H.~Al-Bataineh}	\affiliation{\nmsu}
\author{J.~Alexander}	\affiliation{\stonybrkc}
\author{A.~Al-Jamel}	\affiliation{\nmsu}
\author{K.~Aoki}	\affiliation{\kyoto} \affiliation{\riken}
% ...
\collaboration{PHENIX Collaboration} \noaffiliation

\date{\today}

\begin{abstract}

We present a new analysis of $\jpsi$ production yields in
deuteron-gold collisions at $\sqrt{s_{NN}}$ = 200 GeV using data taken
by the PHENIX experiment in 2003 and previously published in
[S.S. Adler {\it et al.}, Phys. Rev. Lett 96, 012304 (2006)].  The
high statistics proton-proton $\jpsi$ data taken in 2005 is used to
improve the baseline measurement and thus construct updated cold
nuclear matter modification factors ($\rda$).  A suppression of
$\jpsi$ in cold nuclear matter is observed as one goes forward in
rapidity (in the deuteron-going direction), corresponding to a region
more sensitive to initial state low-$x$ gluons in the gold
nucleus. The measured nuclear modification factors are compared to
theoretical calculations of nuclear shadowing to which a $\jpsi$ (or
precursor) breakup cross section is added.  Breakup cross sections
of $\sigma_{breakup} = 2.8^{+1.7}_{-1.4}$ ($2.2^{+1.6}_{-1.5}$) mb are
obtained by fitting these calculations to the data using two different
models of nuclear shadowing.  These breakup cross-section values are
consistent within large uncertainties with the $4.2 \pm 0.5$ mb
determined at lower collision energies.  Projecting this range of cold
nuclear matter effects to copper-copper and gold-gold collisions
reveals that the current constraints are not sufficient to firmly
quantify the additional hot nuclear matter effect.
\end{abstract}

% insert suggested PACS numbers in braces on next line

\maketitle

\section{Introduction}
Understanding the behavior of QCD matter under different conditions of
temperature and density is the subject of intense experimental and
theoretical work in nuclear physics. The transition from hadronic
matter to a quark-gluon plasma at high temperature is expected to be
achieved in high energy heavy ion collisions. The hadronization of
partons in vacuum or cold nuclear matter (i.e. in a nucleus) is also
of keen interest, and represents a nonperturbative and dynamic QCD
process. The formation and interaction of heavy quarkonia (for example
$\jpsi$ mesons) in vacuum, cold nuclear matter and hot nuclear matter present
an excellent laboratory for gaining insights on these
transformations. Recent results from the Relativistic Heavy Ion
Collider reveal a significant suppression of the final $\jpsi$ yield
in central (small impact parameter) $\auau$ reactions at
$\sqrt{s_{NN}}$=200 GeV, relative to expectations scaled from $\pp$
reactions at the same energy~\cite{Adare:2006ns,Adare:2006kf}. A
possible source of this suppression is the screening of the attractive
interaction between the quark-antiquark pair in the hot nuclear
medium, as temperatures are expected to be above the critical
temperature for a quark-gluon plasma transition.  Larger $\jpsi$
suppression is observed at forward rapidity than at midrapidity,
which contradicts models with only color screening of quarkonia
proportional to the local energy density.

Produced $c\bar{c}$ pairs must pass through the remaining nuclear
material from the incident cold nuclei, in addition to surviving any
hot medium environment.  The so-called cold nuclear matter
effects~\cite{Vogt:2004dh}, including modification of initial parton
distribution functions (shadowing, gluon saturation, anti-shadowing,
EMC effect, etc.), initial and final state partonic multiple
scattering, and related initial state parton energy loss need to be
accounted for before firm conclusions can be drawn about the effect of
the hot medium thought to be created.  In fact, these various cold
nuclear matter effects are interesting in their own right, notably in
terms of hadronization time scales, parton energy loss in matter, and
the various initial state effects mentioned above.

This paper presents a new analysis of the modification of $\jpsi$
production in deuteron-gold ($\dau$) collisions relative to
proton-proton ($\pp$) collisions at $\sqrt{s_{NN}}$=200 GeV and the
implications for understanding the $\auau$ and $\cucu$ data at the
same energy. The PHENIX experiment has previously published a result
using $\pp$ and $\dau$ data taken in 2003~\cite{Adler:2005ph}. A
modest $\jpsi$ suppression was observed at forward rapidity (i.e. in
the deuteron moving direction), which is a possible indication of
shadowing of low-$x$ gluons in the gold nucleus. A substantially
larger (more than an order of magnitude) $\pp$ data set was recorded
in 2005 with the $\jpsi$ results published in~\cite{Adare:2006kf}, and
has been used as the baseline for recent $\auau$ and $\cucu$ nuclear
modification factors~\cite{Adare:2006ns,Adare:2007cucu}. The same
$\pp$ data set is used in the analysis presented here to determine the
$\dau$ nuclear modifications more accurately and in a fully consistent
way with those in the $\auau$ and $\cucu$ cases. In addition, during
the two years between the analyses of the 2003 and 2005 data sets
significant improvements in the reconstruction software and signal
extraction method were achieved, as well as an overall better
understanding of the detector performance. These improvements have
been included in this analysis, allowing maximal cancellation of
systematic errors when using the 2005 $\pp$ data sample to form the
$\jpsi$ nuclear modification factor.  We first describe the updated
analysis, then present the new nuclear modification factors and their
implications.

%%%%%%%%%%%%%%%%%%%%%%%%%%%%%%%%%%%%%%%%%%%%%\input{1_Analysis.tex}

%% $Id: 1_Analysis.tex,v 1.23 2007/11/13 21:43:12 mwysocki Exp $

\section{Experiment}

The PHENIX apparatus is described in~\cite{Adcox:2003zm}. It consists
of two sets of spectrometers referred to as the central arms, which
measure particles emitted at midrapidity ($|y|<0.35$), and the muon
arms, measuring particles emitted at backward and forward rapidity
($-2.2 < y < -1.2$ and $1.2 < y < 2.2$).

At midrapidity, $\jpsi$ particles are measured via their decay into
two electrons. Electrons are identified by matching tracks
reconstructed with drift chambers (DC) and pad chambers (PC) to
clusters in the Electromagnetic Calorimeters (EMCAL) and hits in the
Ring Imaging Cerenkov Counters (RICH). In $\dau$ collisions, a charged
track is identified as an electron candidate by requiring at least
three matching RICH phototube hits within a certain radius with
respect to the the center defined by the track projection at the RICH,
a position matching of $\pm 4$~standard deviations between the EMCAL
cluster and the reconstructed track, and a cut on the ratio of energy
to momentum. In $\pp$ collisions the electron identification cuts are
the same except that only two matching RICH phototube hits are
required.

At forward and backward rapidity, $\jpsi$ particles are measured via
their decay into two muons. Muons are identified by matching tracks
measured in Cathode Strip Chambers (referred to as the Muon Tracker,
or MuTR) to hits in alternating planes of Iarocci tubes and steel
absorbers (referred to as the Muon Identifier, or MuID).  Each muon
arm is located behind a thick copper and iron absorber that is meant
to absorb most hadrons produced during the collisions, so that the
measured muons must penetrate 8 to 11 interaction lengths of material
in total.
%For a reconstructed track to be identified as a muon, a
%position matching of 20 (15)~cm between the MuID track and the MuTR
%track is required at positive (negative) rapidity, together with an
%angular matching of 9 degrees.

The $\dau$ data used for this analysis were recorded in 2003 using a
minimum bias trigger that required hits in each of the two beam-beam
counters (BBC) located at positive and negative rapidity
($3<|\eta|<3.9$), and represent integrated luminosities for the
different spectrometers ranging from 1.4 nb$^{-1}$ to 1.7 nb$^{-1}$
(or equivalently 2.7 to 3.4 billion interactions).  This trigger
covers $88 \pm 4$\% of the total $\dau$ inelastic cross section of
2260 mb~\cite{White:2005kp}.  For the electrons, an additional trigger
was used that required one hit above threshold in the EMCAL and a
matching hit in the RICH. For the muons, two additional triggers were
used at different times during the data-taking period.  The muon
triggers are based on information from the MuID, which has five active
detector layers between the steel absorbers.  For the first part of
the data-taking period, one of the tracks was required to reach the
fourth MuID plane, while the other was only required to reach the
second MuID plane.  For the latter part, the trigger required at least
two tracks to reach the fourth MuID plane of Iarocci tubes.

The BBCs are also used to determine the centrality of the $\dau$
collisions by measuring the energy deposited in the counters located
at negative rapidity (in the gold-going direction).  For a given
centrality bin, the average number of equivalent nucleon-nucleon
collisions ($\ncol$) is derived from this energy using a Glauber
calculation~\cite{Miller:2007ri} coupled to a simulation of the
BBC. The centrality bins used in this analysis and the corresponding
number of collisions are listed in Table~\ref{table:centrality}.  To
ensure that the centrality categories are well defined, collisions are
required to be within $\pm30$ cm of the center of the interaction
region.

\begin{table*}
%\centering
\caption{\label{table:centrality}Characterization of the collision
centrality for $\dau$ collisions. First line: centrality bins
used in this analysis; Second line: corresponding number of binary
collisions $\ncol$; lines 3, 4 and 5:
$c={\epsilon^{BBC}_{MB(cent)}}/{\epsilon^{BBC}_{J/\psi}}$ for $\jpsi$ mesons
emitted in the three rapidity ranges used for this analysis.}
\begin{ruledtabular}
\begin{tabular}{c|ccccc}
centrality & 0-20~\%&20-40~\%&40-60~\%&60-88~\%&0-100~\% \\
$\ncol$& $15.4\pm1.0$ &$10.6\pm0.7$&$7.0\pm0.6$&$3.1\pm0.3$&$7.6\pm0.3$\\
$c\;({|y|<0.35})$&$0.95\pm0.03$&$0.99\pm0.01$&$1.03\pm0.01$&$1.04\pm0.03$&$0.94\pm0.02$\\
$c\;({-2.2<y<-1.2})$&$0.93\pm0.03$&$0.99\pm0.01$&$1.04\pm0.01$&$1.05\pm0.03$&$0.94\pm0.02$\\
$c\;({1.2<y<2.2})$&$1.00\pm0.03$&$1.00\pm0.01$&$1.02\pm0.01$&$1.02\pm0.03$&$0.94\pm0.02$\\
\end{tabular}
\end{ruledtabular}
\end{table*}

\section{Signal extraction}
\begin{figure}[htb]
\includegraphics[width=1.0\linewidth]{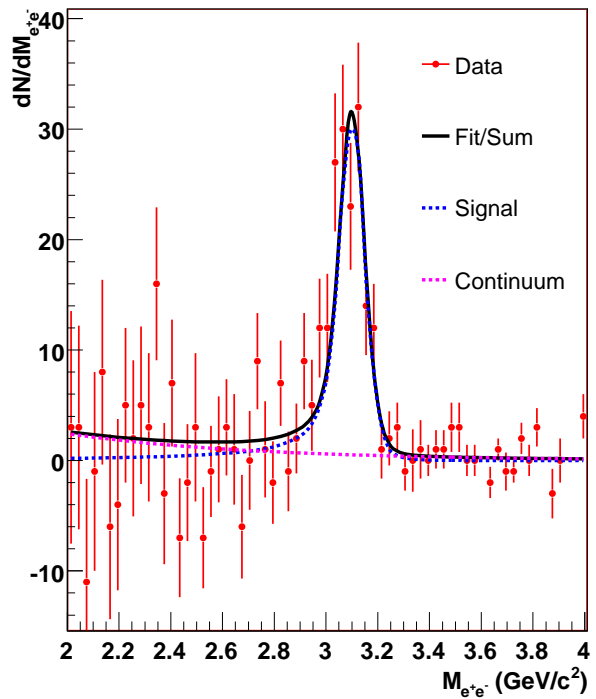}
\caption{\label{fig:mass_plot} (color online) Invariant mass spectra
in minimum bias $\dau$ reactions for $\jpsi \longrightarrow
e^{+}e^{-}$ at $|y|<0.35$, with the functional forms used to extract
the number of reconstructed $\jpsi$ mesons.  }
\end{figure}

The number of $\jpsi$ particles is determined using the invariant mass
distribution of unlike-sign lepton pairs.  At midrapidity, the
$\jpsi$ signal count is obtained via counting the number of
unlike-sign dielectrons after subtracting the like-sign pairs in a
fixed mass window $2.6\leq M_{e^+e^-}\leq3.6$~GeV/c$^2$ or $2.7\leq
M_{e^+e^-}\leq3.5$~GeV/c$^2$, depending on the number of DC hits
required for track reconstruction.  Figure~\ref{fig:mass_plot} shows
the $\jpsi$ mass spectrum after subtracting the background.  The solid
black line is the sum of the $\jpsi$ line shape (dashed
curve) and an exponential function (dot-dashed curve) describing the
continuum component determined from the 2005 $\pp$ data
set~\cite{Adare:2006kf}. The $\jpsi$ line shape function takes into
account the momentum resolution of track reconstruction, internal
radiative effects~\cite{Spiridonov:2004mp}, and external radiative
effects evaluated using a GEANT~\cite{GEANT} simulation of the PHENIX
detector.  The number of $\jpsi$ particles in $\dau$ collisions is too
small to allow a good fit but a comparison between $\dau$ and $\pp$
$\jpsi$ line shapes shows good agreement.  The fraction of $\jpsi$
candidates outside of the mass window due to the radiative effects is
estimated to be $7.2\% \pm 1.0\%$ based on the line shape functions.
The $\jpsi$ signal is also corrected for the dielectron continuum
yield, which originates primarily from open charm and Drell-Yan pairs
inside the mass window.  The estimated contribution is $ 10\% \pm
5\%$, based on the fitting function and PYTHIA~\cite{PYTHIA}
simulations.  Approximately 400 $\jpsi$ mesons are obtained.

\begin{figure}[htb]
\includegraphics[width=1.0\linewidth]{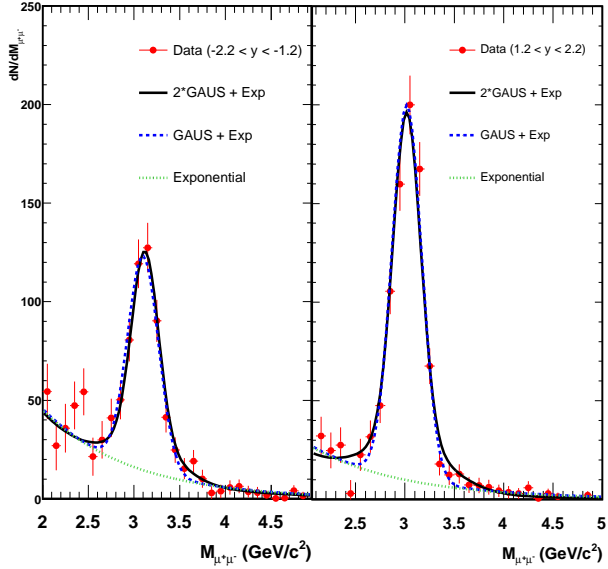}
\caption{\label{fig:mass_plot_dimuon} (color online) Invariant mass
spectra in minimum bias $\dau$ reactions for (left) $\jpsi
\longrightarrow \mu^{+}\mu^{-}$ at $-2.2<y<-1.2$ and (right) $\jpsi
\longrightarrow \mu^{+}\mu^{-}$ at $1.2<y<2.2$, with the functional
forms used to extract the number of reconstructed $\jpsi$ mesons.  }
\end{figure}

At backward and forward rapidity an event mixing technique is now used
to estimate the combinatorial background, whereas the like-sign pairs
were used in the previous analysis~\cite{Adler:2005ph}. The previous
method suffered from a larger statistical uncertainty for bins where
the signal over background is poor. A sample mass distribution after
the subtraction of the mixed event background is shown in
Figure~\ref{fig:mass_plot_dimuon}. Approximately 500 and 750 $\jpsi$
mesons are obtained for backward and forward rapidity, respectively.
The signal counts are determined from this subtracted dimuon
invariant mass distribution with a log-likelihood fit and assuming
three different functional forms and parameters. In all three cases,
an exponential form is used to account for correlated physical
background sources (e.g. Drell-Yan or open charm) and the possible
systematic offset in the normalization of the mixed event
background. The number of $\jpsi$ particles is then estimated either
by direct counting of the remaining number of pairs above the
exponential in the mass range $2.6\leq
M_{\mu^+\mu^-}\leq3.6$~GeV/c$^2$; using a Gaussian function with the
center fixed to the $\jpsi$ mass and the width and integrated yield
left free; or using two Gaussian functions for which both the center
and widths are fixed to the values measured in $\pp$ collisions.  The
two Gaussian functions account for the nonGaussian tails in the
invariant mass distribution.  The normalization of the mixed
background is varied by a systematic uncertainty of $\pm 2$\% prior to
its subtraction from the mass distribution.  This uncertainty is
determined by comparing different normalization methods.  The
corresponding signal variations are included in the systematic
uncertainty. Due to the fit procedure described above, for all $\pp$
and $\dau$ cases this normalization uncertainty results in a very
small systematic uncertainty on the number of measured $\jpsi$
particles. This entire procedure is identical to the one used in
\cite{Adare:2006kf,Adare:2006ns}.

%%%%%%%%%%%%%%%%%%%%%%%%%%%%%%%%%%%%%%%%%%%%%\input{3_Results.tex}

%% $Id: 3_Results.tex,v 1.22 2007/11/13 21:43:14 mwysocki Exp $

\section{Invariant yield}

The $\jpsi$ invariant yield in a given centrality, transverse momentum
and rapidity bin is:
\begin{equation}
\frac{B_{ll}}{2\pi\pt}\frac{d^2N_{\jpsi}}{dp_Tdy}=\frac{1}{2\pi p_T \Delta p_T \Delta y}\frac{N^{\jpsi}_{counts}}{A\epsilon_{rec}\epsilon_{trig} N^{MB}_{evt}} \frac{\epsilon^{BBC}_{MB(cent)}}{\epsilon^{BBC}_{\jpsi}}
\end{equation}
with $B_{ll}$ being the branching ratio for $\jpsi\rightarrow l^+l^-$;
$N^{\jpsi}_{counts}$ the number of reconstructed $\jpsi$ mesons;
$N^{MB}_{evt}$ the number of minimum bias events sampled;
$\epsilon^{BBC}_{MB(cent)}$ the BBC trigger efficiency for minimum
bias events in a given centrality category; $\Delta p_T$ and $\Delta
y$ the $p_T$ and $y$ bin widths; $A$ and $\epsilon_{rec}$ the
acceptance and reconstruction efficiency corrections;
$\epsilon_{trig}$ the additional $\jpsi$ trigger efficiency and
$\epsilon^{BBC}_{\jpsi}$ the BBC efficiency for events containing a
$\jpsi$.  All invariant yields as a function of $p_T$ and $y$
including statistical and systematic uncertainties are given in
Table~\ref{table:ubertable}.

\begingroup \squeezetable
\begin{table*}
\caption{\label{table:ubertable}
The errors quoted in these tables represent 1) the statistical and Type A systematic uncertainties 
added in quadrature, and 2) the Type B systematic uncertainties.  The Type C errors are shown in 
the relevant figures.}
%\centering
\begin{ruledtabular}
\begin{tabular}{ccccc}
\multicolumn{5}{c}{Backward Rapidity Results}\\
Centrality (\%) & $p_{T}$ (GeV/c) & $y$ & Invariant Yield & $R_{dAu}$ \\
\hline
0-100 & All & [-2.2,-1.2] & $(4.264 \pm 0.326 \pm 0.923) \times 10^{-6}$ & $0.90 \pm 0.08 \pm 0.19$ \\
0-100 & All & [-2.2,-1.7] & $(3.583 \pm 0.395 \pm 0.775) \times 10^{-6}$ & $0.95 \pm 0.12 \pm 0.20$ \\
0-100 & All & [-1.7,-1.2] & $(5.292 \pm 0.483 \pm 1.145) \times 10^{-6}$ & $0.90 \pm 0.09 \pm 0.19$ \\
0-100 & 0-1 & [-2.2,-1.2] & $(3.040 \pm 0.460 \pm 0.658) \times 10^{-7}$ & $0.69 \pm 0.11 \pm 0.15$ \\
0-100 & 1-2 & [-2.2,-1.2] & $(1.782 \pm 0.201 \pm 0.386) \times 10^{-7}$ & $0.84 \pm 0.10 \pm 0.18$ \\
0-100 & 2-3 & [-2.2,-1.2] & $(8.141 \pm 0.937 \pm 1.762) \times 10^{-8}$ & $1.44 \pm 0.18 \pm 0.31$ \\
0-100 & 3-4 & [-2.2,-1.2] & $(1.789 \pm 0.359 \pm 0.387) \times 10^{-8}$ & $1.21 \pm 0.26 \pm 0.26$ \\
0-100 & 4-5 & [-2.2,-1.2] & $(4.016 \pm 1.451 \pm 0.869) \times 10^{-9}$ & $1.14 \pm 0.43 \pm 0.24$ \\
 0-20 & All & [-2.2,-1.2] & $(9.084 \pm 0.922 \pm 1.925) \times 10^{-6}$ & $0.94 \pm 0.10 \pm 0.21$ \\
20-40 & All & [-2.2,-1.2] & $(3.676 \pm 0.642 \pm 0.770) \times 10^{-6}$ & $0.55 \pm 0.10 \pm 0.12$ \\
40-60 & All & [-2.2,-1.2] & $(4.013 \pm 0.583 \pm 0.842) \times 10^{-6}$ & $0.92 \pm 0.14 \pm 0.21$ \\
60-88 & All & [-2.2,-1.2] & $(2.062 \pm 0.312 \pm 0.436) \times 10^{-6}$ & $1.07 \pm 0.17 \pm 0.25$ \\
\hline\hline

\multicolumn{5}{c}{ }\\

\hline\hline
\multicolumn{5}{c}{Mid-rapidity Results}\\
Centrality (\%) & $p_{T}$ (GeV/c) & $y$ & Invariant Yield & $R_{dAu}$ \\
\hline
0-100 & All & [-0.35,0.35] & $(6.750 \pm 0.540 \pm 0.950) \times 10^{-6}$ & $0.85 \pm 0.07 \pm 0.15$ \\
0-100 & 0-1 & [-0.35,0.35] & $(6.700 \pm 0.800 \pm 0.940) \times 10^{-7}$ & $1.05 \pm 0.14 \pm 0.21$ \\
0-100 & 1-2 & [-0.35,0.35] & $(2.400 \pm 0.340 \pm 0.340) \times 10^{-7}$ & $0.74 \pm 0.11 \pm 0.15$ \\
0-100 & 2-3 & [-0.35,0.35] & $(1.200 \pm 0.190 \pm 0.170) \times 10^{-7}$ & $0.96 \pm 0.17 \pm 0.19$ \\
0-100 & 3-4 & [-0.35,0.35] & $1.37 \times 10^{-8}$ (90\% CL) & $0.41$ (90\% CL) \\
0-100 & 4-5 & [-0.35,0.35] & $(7.500 \pm 3.600 \pm 1.100) \times 10^{-9}$ & $1.09 \pm 0.61 \pm 0.22$ \\
 0-20 & All & [-0.35,0.35] & $(1.144 \pm 0.160 \pm 0.160) \times 10^{-5}$ & $0.71 \pm 0.10 \pm 0.12$ \\
20-40 & All & [-0.35,0.35] & $(7.990 \pm 1.290 \pm 1.120) \times 10^{-6}$ & $0.71 \pm 0.12 \pm 0.11$ \\
40-60 & All & [-0.35,0.35] & $(6.800 \pm 1.010 \pm 0.950) \times 10^{-6}$ & $0.93 \pm 0.14 \pm 0.14$ \\
60-88 & All & [-0.35,0.35] & $(3.030 \pm 0.500 \pm 0.420) \times 10^{-6}$ & $0.94 \pm 0.16 \pm 0.14$ \\
\hline\hline

\multicolumn{5}{c}{ }\\

\hline\hline
\multicolumn{5}{c}{Forward Rapidity Results}\\
Centrality (\%) & $p_{T}$ (GeV/c) & $y$ & Invariant Yield & $R_{dAu}$ \\
\hline
0-100 & All & [1.2,2.2] & $(3.300 \pm 0.242 \pm 0.592) \times 10^{-6}$ & $0.63 \pm 0.06 \pm 0.11$ \\
0-100 & All & [1.2,1.7] & $(4.522 \pm 0.341 \pm 0.811) \times 10^{-6}$ & $0.68 \pm 0.06 \pm 0.11$ \\
0-100 & All & [1.7,2.2] & $(2.406 \pm 0.224 \pm 0.432) \times 10^{-6}$ & $0.59 \pm 0.06 \pm 0.10$ \\
0-100 & 0-1 & [1.2,2.2] & $(2.779 \pm 0.285 \pm 0.498) \times 10^{-7}$ & $0.55 \pm 0.06 \pm 0.09$ \\
0-100 & 1-2 & [1.2,2.2] & $(1.362 \pm 0.115 \pm 0.244) \times 10^{-7}$ & $0.60 \pm 0.06 \pm 0.10$ \\
0-100 & 2-3 & [1.2,2.2] & $(4.667 \pm 0.566 \pm 0.837) \times 10^{-8}$ & $0.73 \pm 0.10 \pm 0.12$ \\
0-100 & 3-4 & [1.2,2.2] & $(1.472 \pm 0.225 \pm 0.264) \times 10^{-8}$ & $0.93 \pm 0.16 \pm 0.16$ \\
0-100 & 4-5 & [1.2,2.2] & $(2.842 \pm 0.756 \pm 0.510) \times 10^{-9}$ & $0.84 \pm 0.25 \pm 0.14$ \\
 0-20 & All & [1.2,2.2] & $(5.705 \pm 0.501 \pm 0.987) \times 10^{-6}$ & $0.54 \pm 0.05 \pm 0.09$ \\
20-40 & All & [1.2,2.2] & $(4.577 \pm 0.474 \pm 0.783) \times 10^{-6}$ & $0.62 \pm 0.07 \pm 0.11$ \\
40-60 & All & [1.2,2.2] & $(2.950 \pm 0.347 \pm 0.505) \times 10^{-6}$ & $0.62 \pm 0.08 \pm 0.11$ \\
60-88 & All & [1.2,2.2] & $(1.671 \pm 0.195 \pm 0.289) \times 10^{-6}$ & $0.79 \pm 0.10 \pm 0.15$ \\
\end{tabular}
\end{ruledtabular}
\end{table*}
\endgroup

The experiment measures the number of $\jpsi$ particles per BBC
triggered events, which in $\dau$ collisions represent only $88 \pm
4$\% of the total inelastic cross section. An additional correction is
then applied such that the invariant yield represents 100\% of the
total inelastic cross section (as done in previous PHENIX $\dau$
analyses).  The correction factor ratio
$\epsilon^{BBC}_{MB(cent)}/\epsilon^{BBC}_{\jpsi}$ depends {\em a
priori} on the centrality bin and the rapidity range of the measured
$\jpsi$ particles.  The values are given in
Table~\ref{table:centrality}.  The same procedure is applied for $\pp$
collisions, so that the yields are normalized to the $\pp$ total
inelastic cross section of 42 mb.

%This choice is driven by our will to stick the closest to what is measured in the detector. 
%The resulting difference is accounted for when the $\dau$ yield
%is normalized to the number of binary collisions $\ncol$ corresponding
%to 88\% of the total inelastic cross section
%(table~\ref{table:centrality}) when forming the nuclear modification
%factor (see section~\ref{section:rda}).

The acceptance and efficiency corrections are determined using a full
GEANT simulation~\cite{GEANT} of the detector with realistic
resolutions and detector plane efficiencies determined using real
data. Compared to the original result~\cite{Adler:2005ph}, this
simulation benefits from improvements in the understanding of the
detector alignment, resolution, and overall performance. It also
includes the improvements added to the reconstruction software and
used for the recent $\pp$, $\cucu$ and $\auau$
analyses~\cite{Adare:2006kf,Adare:2007cucu,Adare:2006ns}. Although the
additional underlying hit occupancies per event are modest in $\pp$
and $\dau$ collisions, they are accounted for by embedding the
simulated $\jpsi$ mesons in real data events. 
% The observed differences (4-5\%) between embedded and nonembedded events have been included in the systematic uncertainty.
The observed differences (4-5\%) between embedded and nonembedded
events are not significant given the statistics of the simulations,
and therefore are included only as a contribution to the systematic
uncertainty.

The systematic uncertainties in the $\jpsi$ invariant yield
(Table~\ref{tab:syst_error}) are grouped into three categories as in
the previous analyses: point to point uncorrelated (Type A), for which
the points can move independently from one another; point to point
correlated (Type B), for which the points can move coherently though
not necessarily by the same amount; and global uncertainties (Type C),
for which all points move by the same multiplicative
factor. Statistical and uncorrelated systematic uncertainties (Type A)
are summed in quadrature and represented as vertical bars. Type B
uncertainties are represented with boxes. The Type C globally
correlated systematic uncertainties are quoted directly on the
Figures.

\begin{table}
%\centering
\caption{\label{tab:syst_error} 
Sources of systematic uncertainties on the $\jpsi$ invariant yield in
$\dau$ collisions. Columns 2 (3) are the average values at mid
(forward) rapidity. When two values are given, the first (second) is
for peripheral (central) collisions. Uncertainties of type A (type B) are
point to point uncorrelated (correlated).}
\begin{ruledtabular} \begin{tabular}{ccccc}
source&$|y|<0.35$&$|y|\in[1.2,2.2]$&type\\
\hline
%signal extraction & 6.5 to 9~\% & $<$10~\% (not $\pt$ bins) & A\\
signal extraction              & 6~\% & $<$10~\% & A\\
acceptance                     & 8~\% & 10~\% & B\\
efficiency                     & 6~\% & 8 to 20~\% & B\\
run by run variation           & 5~\% & 8~\% & B\\
input $y$, $\pt$ distributions & 2~\% & 4~\% & B\\
embedding                      & 4~\% & 5~\% & B\\
\end{tabular} \end{ruledtabular}
\end{table}

\begin{figure}[htb]
\includegraphics[width=1.0\linewidth]{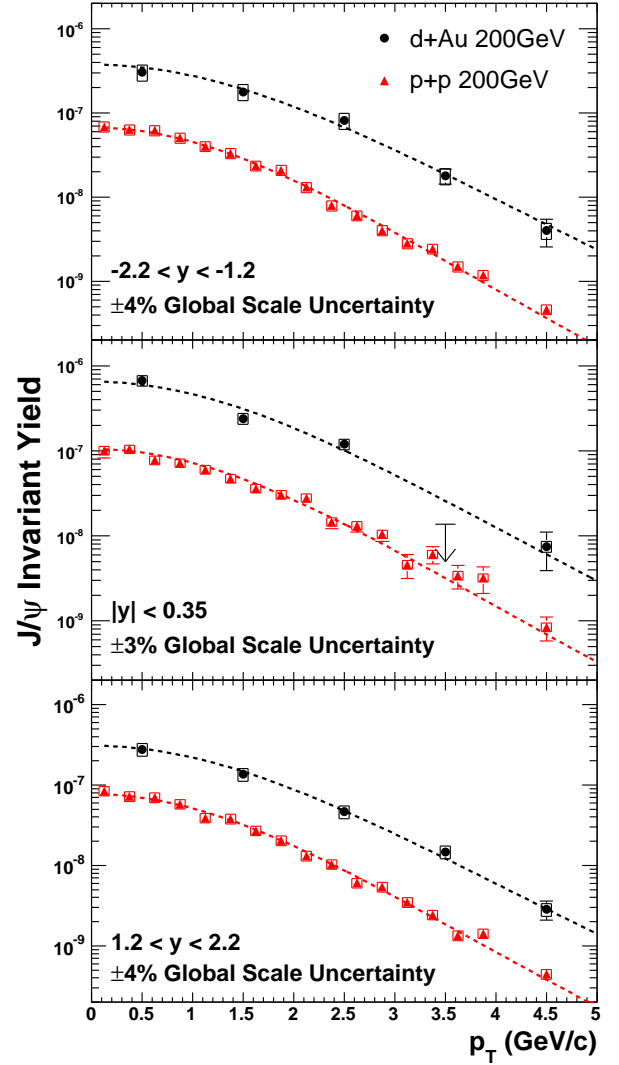}
\caption{\label{fig:pt_spectra} (color on-line) $\jpsi$ invariant
yield versus $\pt$ in $\dau$ collisions and $\pp$ collisions. The
three panels are for rapidity selections $-2.2 < y < -1.2$,
$|y|<0.35$, and $1.2 < y < 2.2$ from top to bottom. See text for
description of the uncertainties and details of the functional fits.
%Type A
%errors are shown as vertical lines. Type B errors are shown as boxes,
%which are typically smaller than the markers. The Type C global scale
%uncertainty error is 10.1\% in the $\pp$ case and XX.Y\% in
%the $\dau$ case. 
}
\end{figure}

Figure~\ref{fig:pt_spectra} shows the invariant $\jpsi$ yield as a
function of transverse momentum for $\dau$ collisions from this
new analysis together with the published invariant yield measured in
$\pp$ collisions~\cite{Adare:2006kf}.
%The systematic
%errors on the $\jpsi$ invariant yield (Table~\ref{tab:syst_error}) are
%grouped into three categories as in previous analyses: point to point
%uncorrelated (Type A) for which the points can move independently from
%one another; point to point correlated (Type B) for which the points
%can move coherently though not necessarily by the same amount; and
%global errors (Type C) for which all points move by the same
%multiplicative factor. Statistical and uncorrelated systematic errors
%are summed in quadrature and represented as vertical bars. Type B
%errors are represented with boxes. 
%The Type C globally correlated systematic errors are 10.1\% for the $\pp$ case and XX.Y\% for
%the $\dau$ case.
%%%%%%%%%%%%%%%%%%%%%%%%%%% 
From these yields, a $\langle\pt^{2}\rangle$ is calculated using the
following generic functional form to fit the data:
\begin{equation}
\frac{d^2N}{p_T dp_T} \sim A(1+(\pt/B)^2)^{-6}
\end{equation}
In order to account for finite $\pt$ binning, the fit function is
first integrated over each $\Delta\pt$ range and the integral is
compared to data in the corresponding bin. The measured
$\langle\pt^{2}\rangle$ as well as the associated statistical and
systematic uncertainties are shown in Table~\ref{table:mean_ptsq}.

\begin{table}[tbh]
\caption{\label{table:mean_ptsq}
%$\langle\pt^2\rangle$ as calculated from the data over the range $0<\pt<5$ GeV/c.
$\langle\pt^2\rangle$ calculated from a fit to the data and restricted
to the range $0<\pt<5$ GeV/c. See text for description of the uncertainties.
%First errors account for the statistical and point-to-point uncorrelated systematic errors on the measured yields. Second errors account for the point-to-point correlated systematic errors
}
\begin{ruledtabular}
\begin{tabular}{ccc}
species & rapidity & $\langle\pt^2\rangle [0,5]$\\
\hline
 d+Au & [-2.2,-1.2] & $4.3 \pm 0.3 \pm 0.4$\\
 d+Au & [-0.35, 0.35] & $3.9 \pm 0.3 \pm 0.3$\\
 d+Au & [1.2,2.2] & $4.0 \pm 0.2 \pm 0.4$\\
\hline
 p+p & [-2.2,-1.2] & $3.4 \pm 0.1 \pm 0.1$\\
 p+p & [-0.35, 0.35] & $4.1 \pm 0.2 \pm 0.1$\\
 p+p & [1.2,2.2] & $3.4 \pm 0.1 \pm 0.1$\\
\end{tabular}
\end{ruledtabular}
\end{table}

In previous $\jpsi$ analyses~\cite{Adare:2006ns,Adare:2006kf}, it was
found that only for the high statistics $\pp$ data set (where the
measurement has good precision out to $\pt \approx 8~GeV/c$) is the
functional form of the $\pt$ spectrum well constrained. In the $\auau$
case, the functional form is not well constrained and leads to a very
large systematic uncertainty on the $\langle\pt^2\rangle$ if
integrated from 0 to $\infty$. The integral was therefore limited to
$\pt<5$~GeV/c, where it is best constrained by the data.
%Therefore in~\cite{Adare:2006ns}, we
%calculate the $\langle\pt^2\rangle$ [0-5 GeV/c] (the same quantity
%but only restricted to the range 0 to 5 GeV/c, where it is best
%constrained by the data). As a result this quantity has a smaller
%systematic from the variation in functional form.
The $\dau$ data set suffers from the same statistical
limitations and the same truncation to $\pt<5$~GeV/c is
applied. Finally, this constraint is also applied to the $\pp$
case to make a direct comparison possible.
% although the large statistics of this sample allow for a fully integrated measurement (that is, from 0 to $\infty$).

Two uncertainties are quoted in Table~\ref{table:mean_ptsq}. The first
corresponds to the statistical and point-to-point uncorrelated
systematic uncertainties (Type A) on the measured yields. It is
obtained directly from the fit using the second derivatives of the
$\chi^2$ surface at the minimum. The second corresponds to the
point-to-point correlated systematic uncertainties (Type B).
% and could not be easily
%handled in the $\chi^2$ minimization because the magnitude (and the sign) of the
%correlation is largely unknown. 
The contribution from the Type B uncertainty is estimated
independently by coherently moving the measured points within the one
standard deviation limit given by these uncertainties, allowing them
to be either correlated or anti-correlated, and then re-doing the fit
in all cases.  The largest difference observed in the values obtained
by the fit is used as an upper limit to the 1-sigma point-to-point
correlated uncertainties on the $\langle\pt^2\rangle$.

% calculate the restricted $\langle\pt^2\rangle$
%[0-5 GeV/c], and also for the $\pp$ case to be able to make a
%direct comparison. 

In the previous publication~\cite{Adler:2005ph}, values for the fully
integrated $\langle\pt^{2}\rangle$ in $\pp$ and $\dau$ are quoted. A
significant systematic uncertainty originating from not knowing the
functional form to best describe the data was found since then that
was not included in the uncertainty quoted in the paper. In addition,
the new analysis revealed a bias in the previous result that increased
the signal, particularly in the lowest $\pt$ bin. This bias is now
corrected by using the mixed event background subtraction technique
described above together with the modified log-likelihood fit over a
more appropriate range, corresponding to the region where the physical
background can accurately be described by a single exponential
function. Finally, no separate treatment of the point-to-point
correlated systematic uncertainties was performed at that time, since
it was assumed that it would move all points in the same direction
(positive correlation) and thus have no impact on the measured
$\langle\pt^{2}\rangle$.

The data, within uncertainties, includes the possiblity of a modest
broadening of the transverse momentum distribution relative to $\pp$
collisions. This is often attributed to initial and final state
multiple scattering, sometimes referred to as the ``Cronin effect.''
However, in calculating the $\Delta\langle\pt^2\rangle = {\langle
\pt^2\rangle_{dAu}-\langle\pt^2\rangle_{pp}}$ one finds this effect
needs reduced uncertainties from future larger data sets to make any
firm conclusions.

\begin{figure}[htb]
\includegraphics[width=1.0\linewidth]{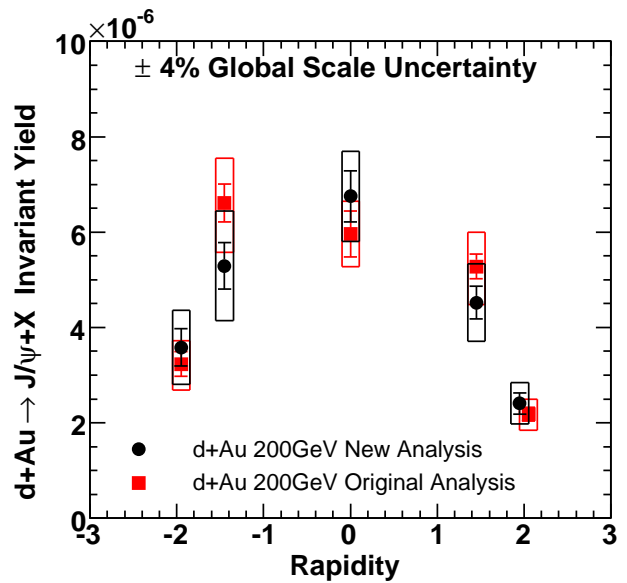}
\caption{\label{fig:dndy_dau} (color online) $\jpsi$ invariant yield
as a function of rapidity for $\dau$ collisions. Shown are the
new analysis results from this paper, in addition to the originally
published results~\cite{Adler:2005ph} using the same data.  
The global systematic uncertainty quoted is for the new analysis.}
\end{figure}

% insert text here on dN/dy figures
Figure~\ref{fig:dndy_dau} shows the $\jpsi$ invariant yield,
integrated over all $\pt$, as a function of rapidity for $\dau$
collisions.  Shown are the results of the new analysis presented in
this paper, as well as the previously published
results~\cite{Adler:2005ph} using the same data set.  Overall the
agreement of the two analysis results is good.  The two sets of points
differ in the reconstruction software, analysis cuts, and signal
extraction technique. Thus many of the systematic uncertainties are
different, and even the statistical uncertainties are not identical
due to the different analysis cuts and the use of event mixing to
estimate the combinatorial background in the new analysis, as opposed
to the like-sign mass distribution used in~\cite{Adler:2005ph}.

\begin{figure}[htb]
\includegraphics[width=1.0\linewidth]{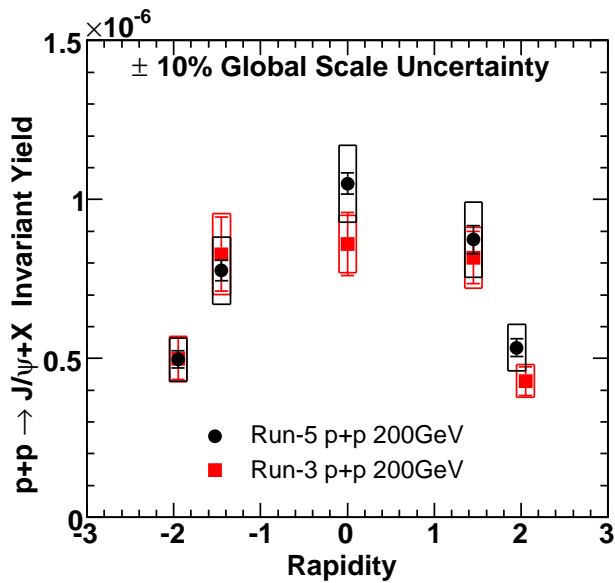}
\caption{\label{fig:dndy_pp} (color online) $\jpsi$ invariant yield as
a function of rapidity for $\pp$ collisions.  Shown are the
high statistics results from 2005 $\pp$ PHENIX data taking
period~\cite{Adare:2006kf}, and the originally published
results~\cite{Adler:2005ph} using the 2003 $\pp$ data set.
The global systematic uncertainty quoted is for the new analysis.}
\end{figure}

Figure~\ref{fig:dndy_pp} shows the $\jpsi$ invariant yield for $\pp$
collisions, from both the published high statistics result from
Run-5~\cite{Adare:2006kf}, as well as the lower statistics result from
Run-3 as published in~\cite{Adler:2005ph}. In both cases the points
are in good agreement within the systematic uncertainty bands. A new
analysis of the Run-3 $\pp$ lower statistics data set was also
performed using the same technique and analysis cuts as for $\dau$
collisions. It also shows good agreement with these two sets of
measurements, albeit with larger statistical uncertainties.
% but is not included
%here due to the smaller statistical sample.

In Figures~\ref{fig:dndy_dau} and~\ref{fig:dndy_pp}, the highest
rapidity point is not located exactly at the same rapidity position
between the original and the new analysis. This is due to the fact
that the positive rapidity muon arm has a slightly larger rapidity
coverage than the negative rapidity arm. This property was used in the
2003 analysis to include additional $\jpsi$ mesons at forward rapidity
in order to probe a slightly lower region of $x$.  It was found,
however, that there were very few counts in this region and that the
asymmetric rapidity range created additional difficulties when
comparing the results measured at forward and backward rapidity (in
case of symmetric collisions) and when comparing the results obtained
in $\pp$ collisions to $\cucu$ or $\auau$ collisions, for which this
extra rapidity coverage was not available (due to high occupancy
limitations at forward rapidity). As a consequence, it was decided for
the later analyses to forgo the extra few $\jpsi$ counts at very
forward rapidity and use the same width rapidity bins at both positive
and negative rapidity.

% end discussion on changes in the data points 

\section{\label{section:rda}Nuclear Modification Factor}

The $\jpsi$ nuclear modification factor in a given centrality and
rapidity bin is:
\begin{equation}
\rda = \frac{1}{\langle \ncol\rangle} \frac{dN_{\jpsi}^{d+Au} / dy}{dN_{\jpsi}^{p+p} / dy}
\end{equation}
with $dN_{\jpsi}^{\rm dAu} / dy$ being the $\jpsi$ invariant yield
measured in $\dau$ collisions; $dN_{\jpsi}^{\rm pp} / dy$ the $\jpsi$
invariant yield measured in $\pp$ collisions for the same rapidity bin
and $\langle\ncol\rangle$ the average number of binary collisions in
the centrality bin under consideration, as listed in
Table~\ref{table:centrality}.  All $\rda$ values as a function of
$p_T$, $y$ and centrality including statistical and systematic
uncertainties are given in the Appendix Table~\ref{table:ubertable}.
%$\langle\ncol\rangle$ is derived from the energy
%deposited in the negative rapidity BBC using a Glauber calculation
%coupled to a simulation of the BBC.

\begin{figure}[htb]
\includegraphics[width=1.0\linewidth]{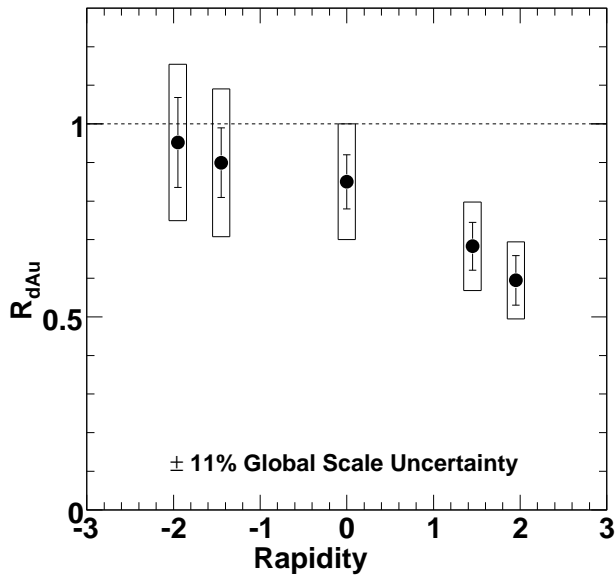}
\caption{\label{fig:rdau_vs_y} $\jpsi$ nuclear modification factor
$\rda$ as a function of rapidity.}
\end{figure}

Figure~\ref{fig:rdau_vs_y} shows the nuclear modification factor
$\rda$ calculated using the $\dau$ new analysis presented in this
paper for the numerator and the 2005 $\pp$ data for the denominator.
In contrast to the previous analysis~\cite{Adler:2005ph}, where the
$\pp$ results were symmetrized around $y=0$ before calculating $\rda$
to compensate for lower $\jpsi$ statistics in the 2003 $\pp$ data set,
in this case the $\rda$ values are calculated independently at each
rapidity.
%Note that the $\rda$ values are calculated point by
%point to maximize the cancellation of systematic uncertainties.  
%This is different from the previous analysis~\cite{Adler:2005ph} where the
%$\pp$ results were symmetrized around $y=0$ before calculating $\rda$
%to compensate for lower $\jpsi$ statistics in the 2003 $\pp$ data set.

%Since identical analysis and signal extraction
%techniques are used together with similar selection cuts, there is
%significant cancellation between the systematic uncertainties when forming
%the ratio, notably concerning the uncertainty on the knowledge of the
%detector geometrical acceptance. Additionally, as can be seen in
%Figure~\ref{fig:dndy_pp} the statistical and systematic uncertainties
%associated with the $\pp$ result are smaller than those
%of 2003 data set. On the other hand some systematic uncertainties have been
%revised (and increased) for the $\dau$ points with respect to
%the original analysis.

The understanding of the detector performance in terms of alignment,
resolution and efficiency has significantly improved between this
analysis and previously published PHENIX $\dau$
results~\cite{Adler:2005ph}. This resulted in changes in the
reconstruction software, analysis cuts, signal extraction technique
and handling of both the physical and combinatorial background in the
dilepton invariant mass distribution.
%Applying these changes to the $\dau$ data set leads to an overall decrease of the estimated
%$\jpsi$ yield.  
Simultaneously, the systematic uncertainties associated with the
measurement have also been re-evaluated in a way consistent with what
was learned for the $\pp$, $\cucu$, and $\auau$ analyses. The new
uncertainties are in general larger, although some of them cancel with
their $\pp$ counterpart when forming $\rda$. This approximately
counterbalances the reduction of the statistical uncertainty achieved
by using the 2005 $\pp$ data set as a reference. Additionally, the
$\jpsi$ production cross sections in $\pp$ collisions measured in
2005~\cite{Adare:2006kf} are compatible within uncertainties, but
higher than the values used in~\cite{Adler:2005ph} (based on the 2003
$\pp$ data set) by about 13~\%. As a consequence, the new nuclear
modification factors are systematically lower than the ones previously
published by about 5 to 20\% for most points, depending on the $\pt$,
$y$ or centrality bin that is considered.

%This is notably the case for the error associated to the MuID efficiency.

Within uncertainties, the nuclear modification factors are consistent
with $\rda = 1.0$ at negative and midrapidities, and are
significantly lower than 1.0 at forward rapidity only, that is, in the
deuteron-going direction. This trend is similar to that shown in
Figure 1 of~\cite{Adler:2005ph}, although the new values are
systematically smaller for all rapidity bins.

\begin{figure}[htb]
\includegraphics[width=1.0\linewidth]{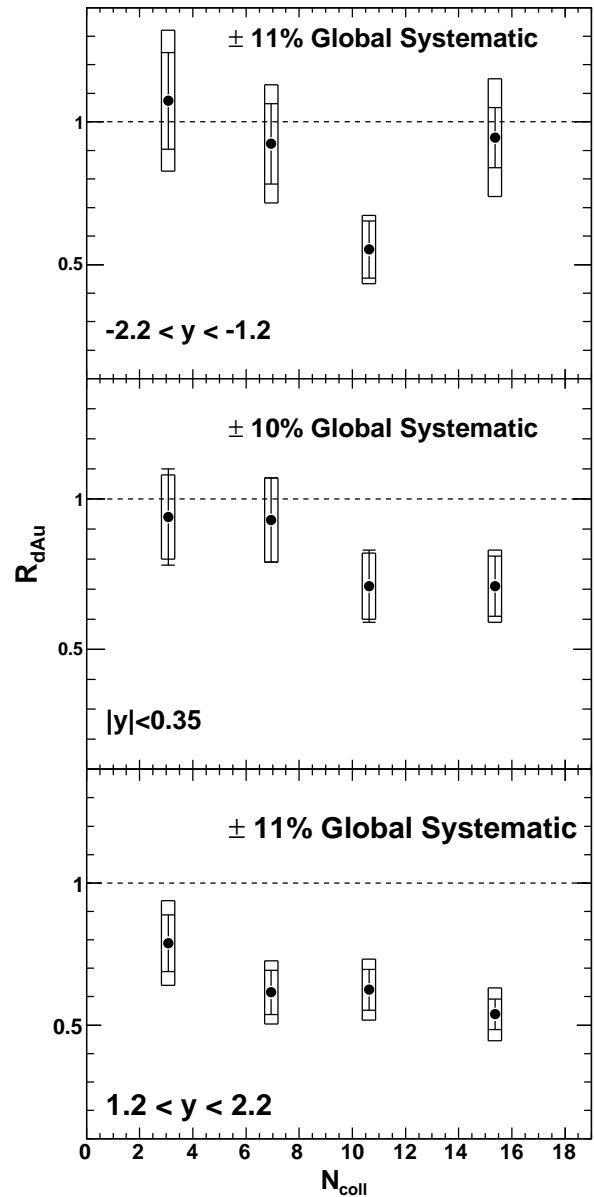}
\caption{\label{fig:rdau_vs_ncoll} $\jpsi$ nuclear modification factor
$\rda$ as a function of $\ncol$ for three rapidity ranges.}
\end{figure}

Figure~\ref{fig:rdau_vs_ncoll} shows the $\jpsi$ nuclear modification
factor in $\dau$ collisions as a function of the number of binary
collisions for three rapidity ranges and four centrality classes.
Only at forward rapidity is there statistically significant suppression.
%, and the current statistical and systematic uncertainties do not
%allow one to constrain the centrality dependence.
%It is notable that there is significant overlap between the different $\ncol$ bins
%due to the limited correlation between $\ncol$ and the energy measured
%in the BBC at positive pseudo-rapidity (as discussed later and shown
%in Figure~\ref{fig:nagle_glauber}).
%This effect must be carefully accounted for in theoretical model calculations prior to any
%comparison to our data .

%%%%%%%%%%%%%%%%%%%%%%%%%%%%%%%%%%%%%%%%%%%%%\input{4_Discussion.tex}

%% $Id: 4_Discussion.tex,v 1.25 2007/11/13 21:43:16 mwysocki Exp $

\section{Discussion}
As stated in the Introduction, the $\dau$ data is interesting both to
fundamentally understand issues of quarkonia and cold nuclear matter
and also to separate these effects from hot nuclear matter effects in
heavy ion collisions.  In order to address both issues, we compare the
experimental data with two different models including both
modification of the initial parton distribution functions (PDF) and a
free parameter to account for the breakup of correlated $c\bar{c}$
pairs that might have otherwise formed $\jpsi$ mesons.  Note that
often in the literature, this breakup process in cold nuclear matter
is referred to as an absorption cross section of the $\jpsi$ particles
on the nucleons in the nucleus.  Here we avoid this nomenclature both
because the object that is ``absorbed'' is generally not a
fully-formed $\jpsi$ but rather a $c\bar{c}$ pair, and because the
actual process is more a breakup of this pair, rather than the
absorption of it.

\begin{figure}[htb]
\includegraphics[width=1.0\linewidth]{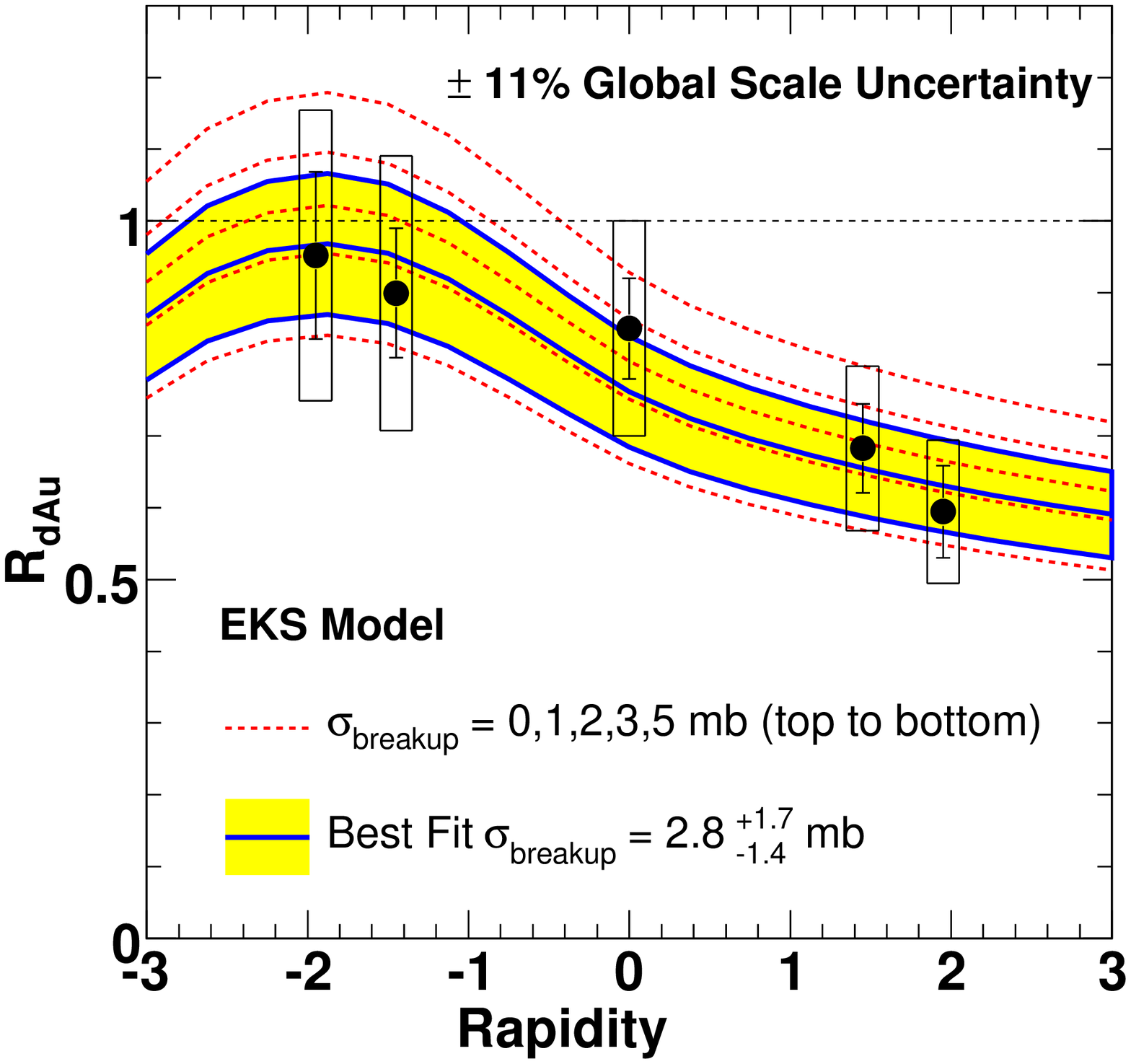}
\includegraphics[width=1.0\linewidth]{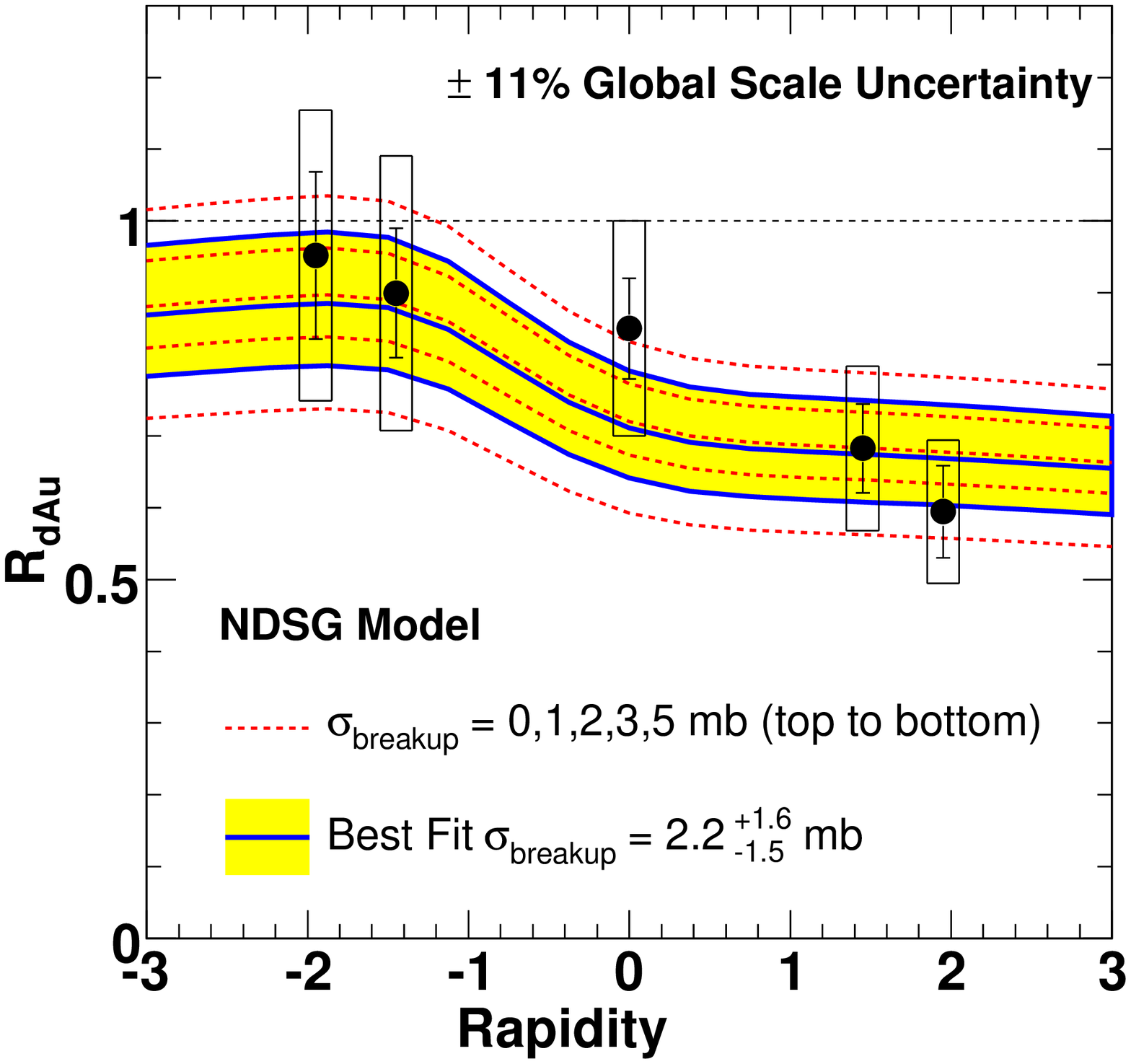}
\caption{ (color online) $\rda$ data compared to various
theoretical curves for different $\sigma_{breakup}$ values.  Also,
shown as a band are the range of $\sigma_{breakup}$ found to be
consistent with the data within one standard deviation.  The top panel
is a comparison for EKS shadowing~\cite{Eskola:2001ek}, while the
bottom panel is for NDSG shadowing~\cite{NDSG}.}
\label{fig:model_comparisons_dau}
\end{figure}

Shown in Figure~\ref{fig:model_comparisons_dau} is the nuclear
modification factor $\rda$ as a function of rapidity in comparison to
theoretical calculations~\cite{Vogt:2004dh} that include either
EKS~\cite{Eskola:2001ek} or NDSG~\cite{NDSG} shadowing models for the
nuclear PDFs.  In each case an additional suppression associated with
a $\sigma_{breakup}$ is also included.  Note that there is no \emph{ab
initio} calculation of this cross section, and while one might expect
a similar value to results at lower energy~\cite{Alessandro:2006jt},
it need not be identical.

\begin{figure}[htb]
\includegraphics[width=1.0\linewidth]{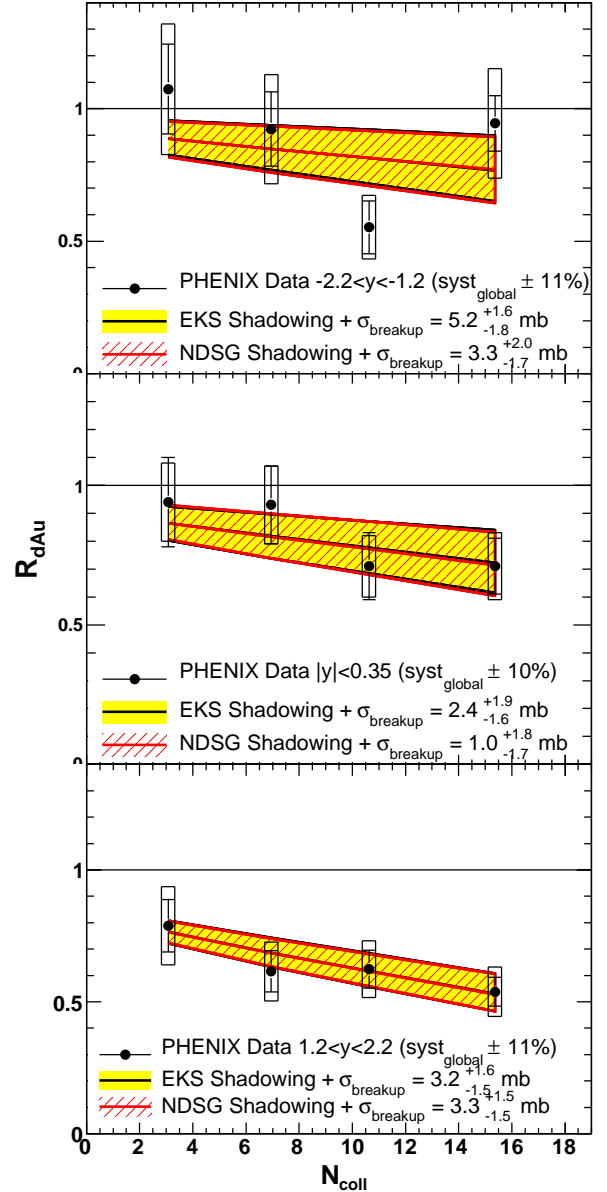}
\caption{ (color online) $\rda$ data as a function of $\ncol$ for
three different rapidity ranges.  Overlayed are theoretical curves representing
the best fit $\sigma_{breakup}$ values  as determined in each rapidity range 
separately, utilizing EKS and NDSG nuclear PDFs and a simple
geometric dependence.   Also, shown as bands are
the range of $\sigma_{breakup}$ found to be consistent with the data
within one standard deviation.}
\label{fig:model_comparisons_dau_ncoll}
\end{figure}

Taking full account of the statistical and systematic uncertainties on
the experimental data, the breakup cross section is determined under
certain assumptions.  We have followed the statistical procedure
detailed in~\cite{phenix_qhat_constraint_ppg079}.  If we assume that
the EKS modified nuclear PDFs are exactly correct, and that the only
additional suppression is accounted for by $\sigma_{breakup}$, then
the data constrains $\sigma_{breakup} = 2.8 ^{+1.7}_{-1.4}$ mb with
the uncertainties as one standard deviation. Similarly, if we assume
the NDSG modified nuclear PDFs, then we obtain $\sigma_{breakup} = 2.2
^{+1.6}_{-1.5}$ mb.  These breakup cross-section values are
consistent (within the large uncertainties) with the $4.2 \pm 0.5$ mb
determined at lower energies at the CERN-SPS~\cite{Alessandro:2006jt}.
The extracted breakup cross section at lower energies assumes no
contribution from the modification of nuclear PDFs.  At the lower
energies, $\jpsi$ production is sensitive to higher-$x$ partons
in the anti-shadowing regime where the modifications are expected to
be smaller and in the opposite direction~\cite{Lourenco:2006sr}.

\begin{table}[thb]
%\centering
\caption{\label{tab:sigma_breakup} Most probable values and one
standard deviations of $\sigma_{breakup}$ assuming two different
shadowing models, from a fit to minimum bias $\rda$ points as a
function of rapidity (Figure~\ref{fig:model_comparisons_dau}), and
fits to $\rda$ as a function of $\ncol$ in three separate rapidity
bins (Figure~\ref{fig:model_comparisons_dau_ncoll}).}
\begin{ruledtabular} \begin{tabular}{ccc}
Fit Range in $y$ & EKS (mb) & NDSG (mb) \\
\hline
All & $2.8^{+1.7}_{-1.4}$ & $2.2^{+1.6}_{-1.5}$ \\
$[-2.2,-1.2]$ & $5.2^{+1.6}_{-1.8}$ & $3.3^{+2.0}_{-1.7}$ \\
$[-0.35,0.35]$ & $2.4^{+1.9}_{-1.6}$ & $1.0^{+1.8}_{-1.7}$ \\
$[1.2,2.2]$ & $3.2^{+1.6}_{-1.5}$ & $3.3^{+1.5}_{-1.5}$ \\
\end{tabular}
\end{ruledtabular}
\end{table}

The modified nuclear PDFs from EKS and NDSG are constrained from other
experimental measurements such as deep inelastic scattering from
various nuclear targets and the resulting $F_2(A)$ structure
functions.  A geometric parametrization of these PDFs based on the
path of the parton through the nucleus is described
in~\cite{Klein:2003dj} and~\cite{Vogt:2004dh}.
%This geometric dependence is not constrained by experimental data.  
One can test this geometric dependence by comparison with the $\dau$
nuclear modification factors as a function of $\ncol$.  Using this
geometric dependence, the most probable $\sigma_{breakup}$ is
calculated independently in three rapidity ranges (see
Table~\ref{tab:sigma_breakup}).  The corresponding nuclear
modification values and their one standard deviation bands are shown
as a function of $\ncol$ in
Figure~\ref{fig:model_comparisons_dau_ncoll}.  The two calculations
with EKS and NDSG nuclear PDFs yield almost identical bands since the
same geometric dependence is used in both cases.  However, each band
represents a different balance of modification due to the nuclear PDF
and the breakup cross section.

\begin{figure}[thb]
\includegraphics[width=1.0\linewidth,clip=]{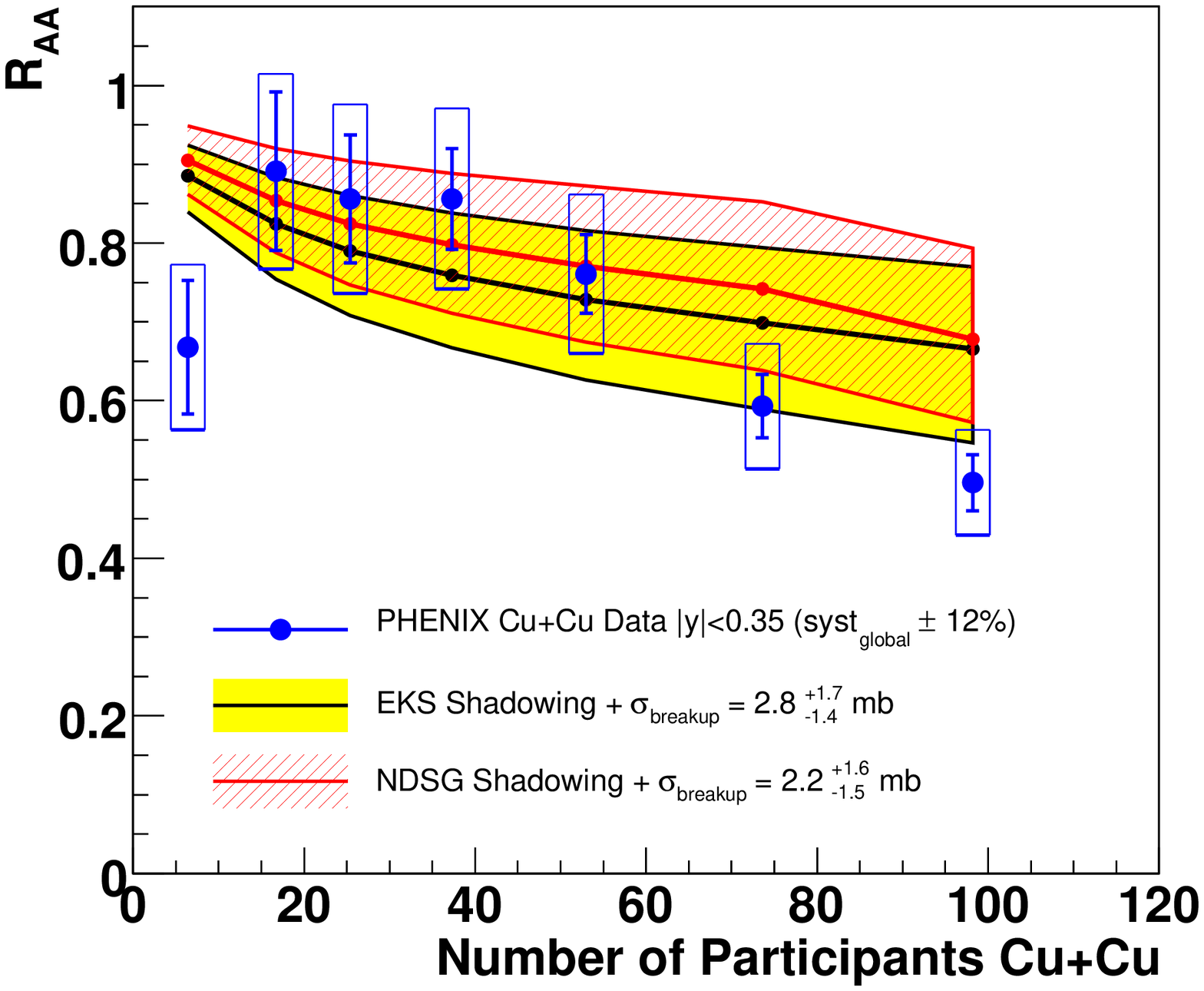}
\includegraphics[width=1.0\linewidth,clip=]{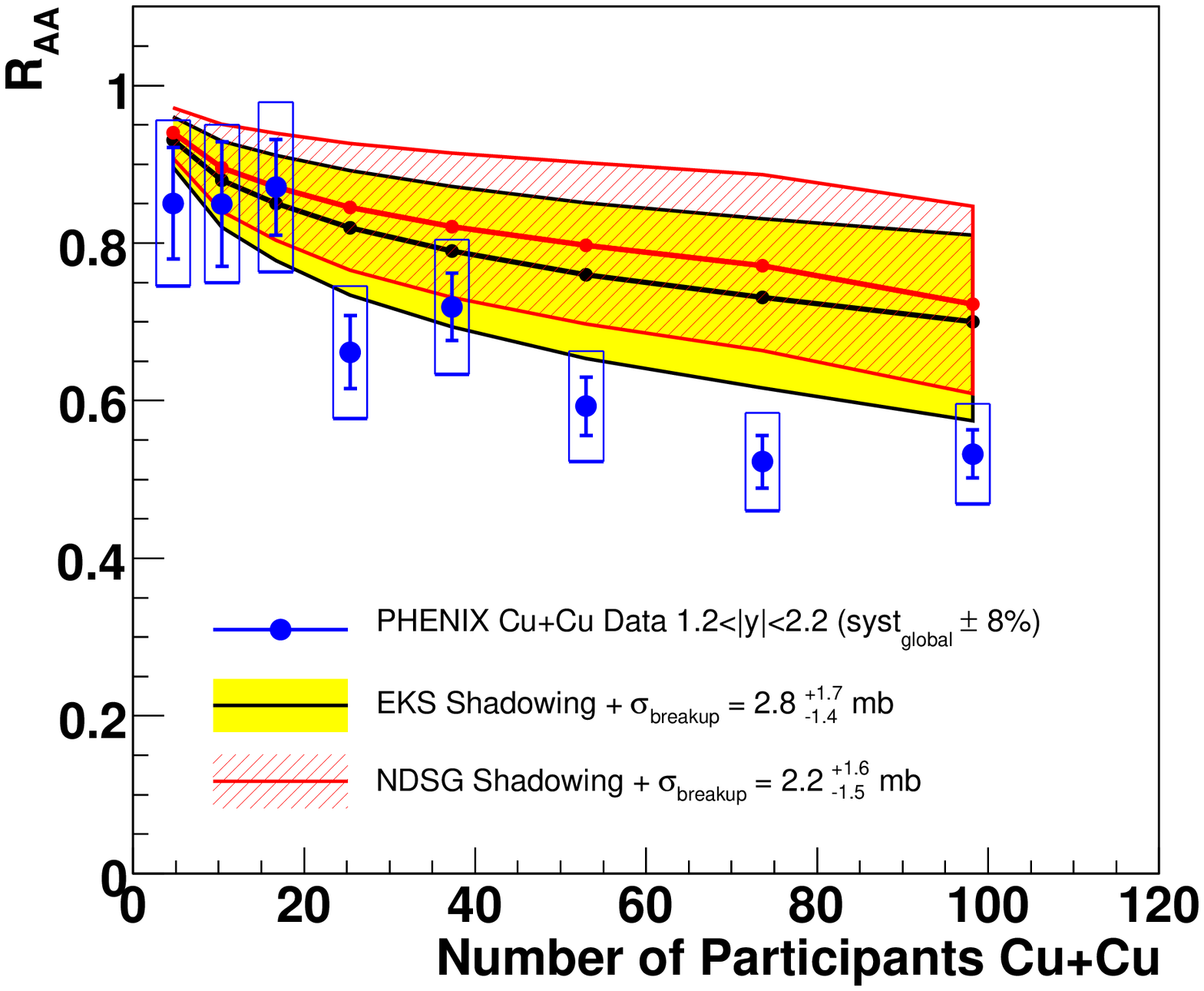}
\caption{(color online) $R_{AA}$ for Cu+Cu~\cite{Adare:2007cucu}
collisions compared to a band of theoretical curves for the
$\sigma_{breakup}$ values found to be consistent with the $\dau$ data
as shown in Figure~\ref{fig:model_comparisons_dau}. The top figure
includes both EKS shadowing~\cite{Eskola:2001ek} and NDSG
shadowing~\cite{NDSG} at midrapidity.  The bottom figure is the same
at forward rapidity.}
\label{fig:model_comparisons_cucu} 
\end{figure}

\begin{figure}[thb]
\includegraphics[width=1.0\linewidth,clip=]{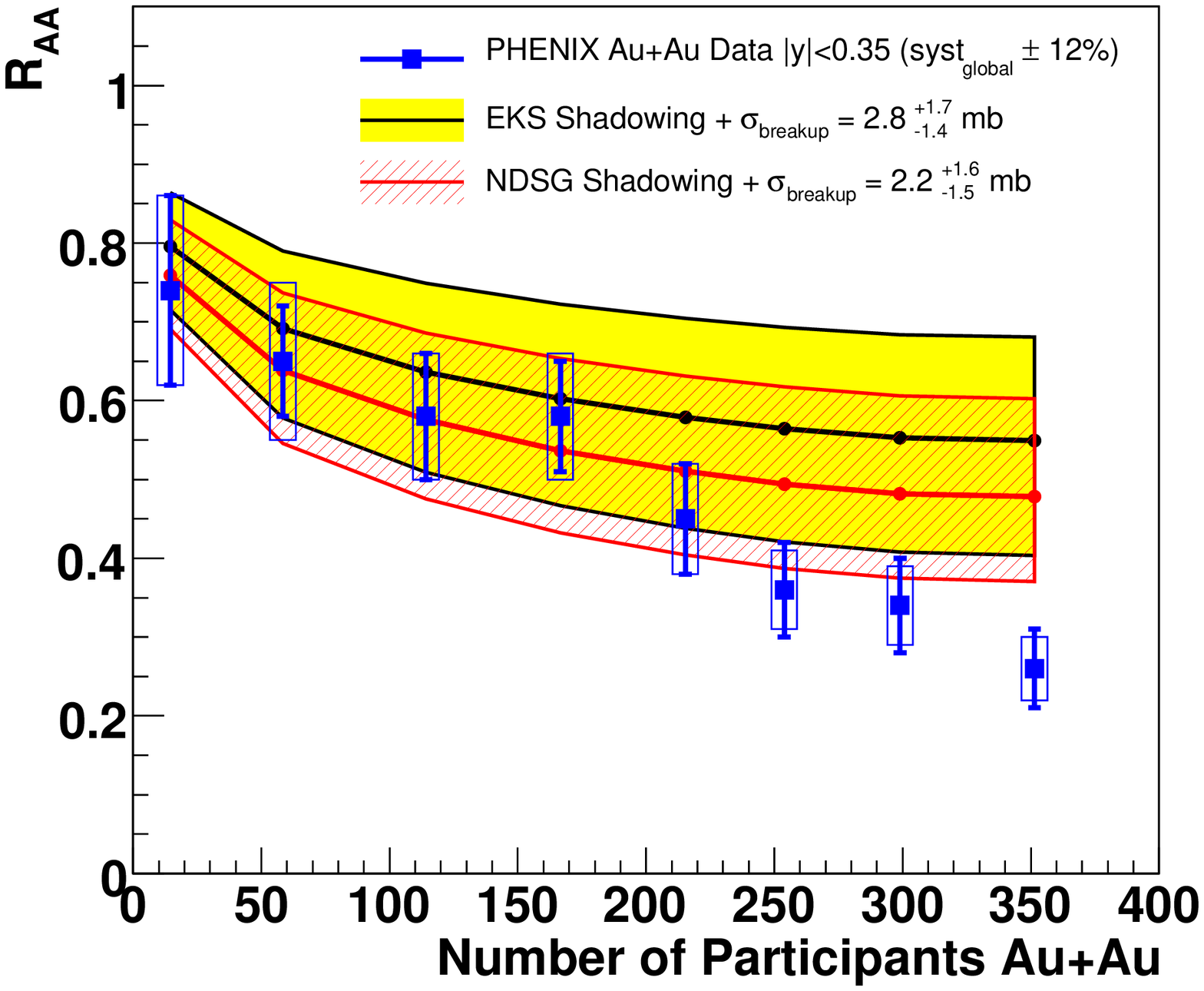}
\includegraphics[width=1.0\linewidth,clip=]{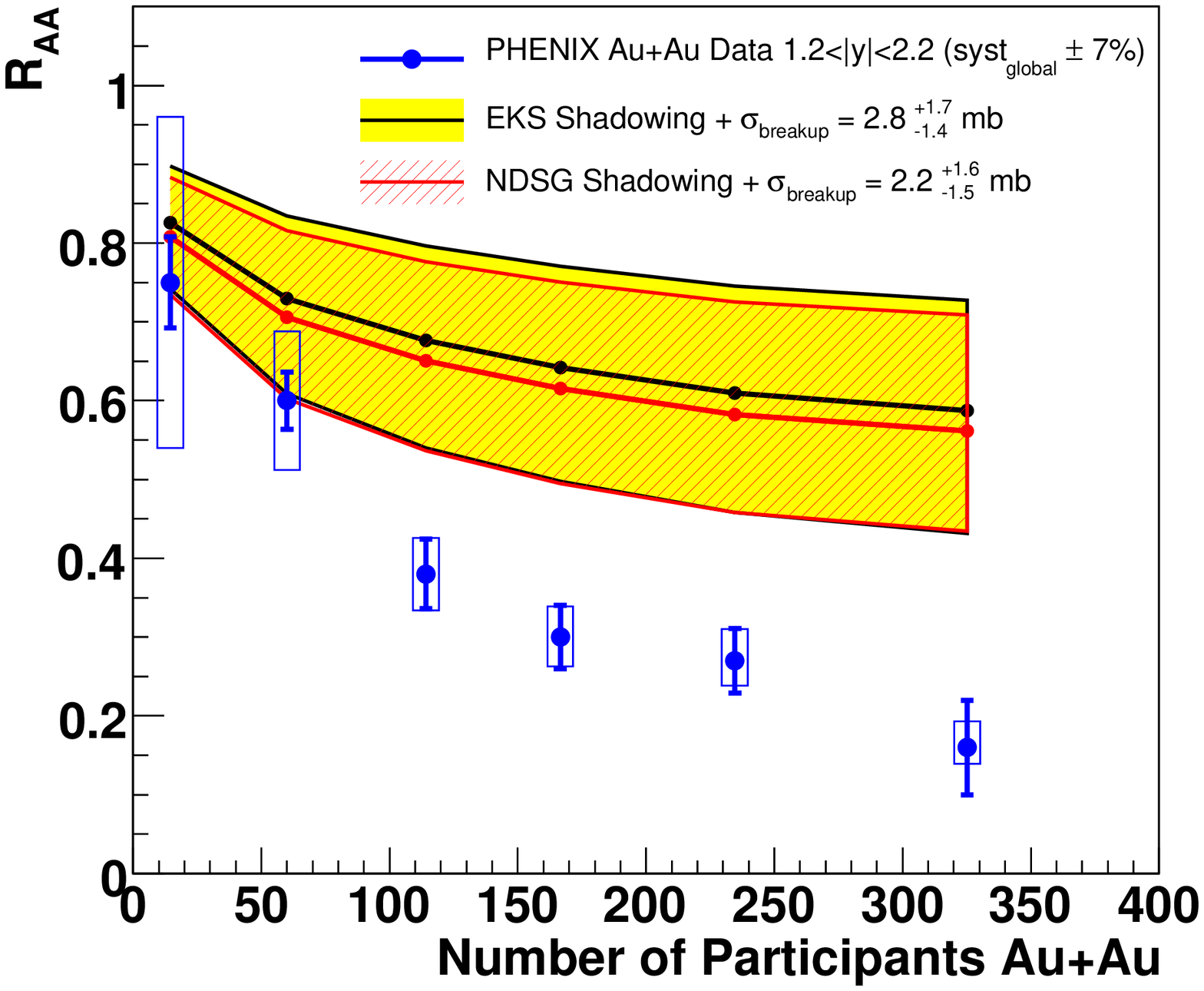}
\caption{(color online) $R_{AA}$ for Au+Au~\cite{Adare:2006ns}
collisions compared to a band of theoretical curves for the
$\sigma_{breakup}$ values found to be consistent with the $\dau$ data
as shown in Figure~\ref{fig:model_comparisons_dau}. The top figure
includes both EKS shadowing~\cite{Eskola:2001ek} and NDSG
shadowing~\cite{NDSG} at midrapidity.  The bottom figure is the same
at forward rapidity.}
\label{fig:model_comparisons_auau} 
\end{figure}

For both the EKS and the NDSG PDFs, the values of $\sigma_{breakup}$
extracted from the overall rapidity dependence of $\rda$ and from the
$\ncol$ dependence of $\rda$ within the different rapidity ranges are
consistent within the large systematic uncertainties.  It should be
noted that though the $1-$p-value for the best fit is poor at backward
rapidity (as can be seen in Fig.~\ref{fig:model_comparisons_dau_ncoll}), 
there is still a well defined maximum in the likelihood function for 
$\sigma_{breakup}$.  A future higher precision $\rda$ measurement as a 
function of centrality will be crucial to constrain the exact geometric 
dependence.

One can also utilize this model to do a consistent calculation of the
contribution from cold nuclear matter effects that should be present
in $\cucu$ and $\auau$ collisions. These contributions, using the best
fit value of $\sigma_{breakup}$ and their one standard deviation
values extracted from the data in
Figure~\ref{fig:model_comparisons_dau} for each of the two shadowing
models, are shown in Figures~\ref{fig:model_comparisons_cucu}
and~\ref{fig:model_comparisons_auau}.  In the $\cucu$ case, $\jpsi$
production is not suppressed beyond cold nuclear matter effects at
midrapidity or at forward rapidity, within the limits of the large
error bands, and the midrapidity data in the $\auau$ case is
similarly inconclusive.  However, there is a significant suppression
in the data at forward rapidity, beyond the uncertainties in both the
data and the projection.  It should be noted that the uncertainty band
at forward and midrapidities are entirely correlated, as they reflect
only the uncertainty in the $\sigma_{breakup}$.  There is no
systematic uncertainty included for the choice of modified nuclear PDF
model, which is the only way to change the relative suppression
between forward and midrapidities within the context of this
calculation.  The more data-driven calculation described later in this
section, however, is performed independently at different rapidities
and does not suffer the same stipulation.

It should also be noted that the theoretical calculations yield
$R_{AA}$ as a continuous function of the number of participants,
whereas the data points are at discrete values representing a
convolution of the modification factor with the $N_{part}$
distribution within a particular centrality category.  A Glauber
simulation combined with a Monte Carlo of the PHENIX experimental
trigger and centrality selection is utilized to convert the continuous
theory predictions into discrete predictions in the simulated PHENIX
centrality categories.  Thus, the results shown in the figures are in
fact predictions for the matched event selection categories of the
experimental data points.

In order to explore the cold nuclear matter constraints further, an
alternative data-driven method proposed
in~\cite{GranierdeCassagnac:2007aj} is used.  This approach assumes
that there is a single modification factor parameterizing all cold
nuclear matter effects that is a simple function of the radial
position in the nucleus.  This computation has the advantage of not
having to assume a specific shadowing scheme and a specific breakup
cross section, but instead relies only on the measured impact
parameter dependence.  It assumes that the cold nuclear matter effects
suffered by a $\jpsi$ in a $\auau$ collision at a given rapidity are
the product of the modifications measured in $\dau$ collisions at the
same rapidity and the modifications measured at the opposite rapidity
(or equivalently in a $\rm{Au+d}$ collision).  This assumption holds
for the two effects considered so far, namely shadowing and subsequent
breakup. It also assumes that the same parton distributions are
sampled by the $\jpsi$ particles observed in the (wide) rapidity range
in $\auau$ and $\dau$ collisions.
%sample the same parton distributions.  
Note that since this model implicitly includes any possible modified
nuclear PDFs, the modification factors may have an $x$-dependence
that is accounted for by considering the backward, mid and forward
rapidity $\dau$ data.
The different rapidity regions are sensitive to the initial-state
partons in the gold nucleus in three broad ranges of $x$,
corresponding to $x \approx$ 0.002-0.01, 0.01-0.05, and 0.05-0.2, as
determined from PYTHIA.

\begin{figure}[htb]
\includegraphics[width=1.0\linewidth]{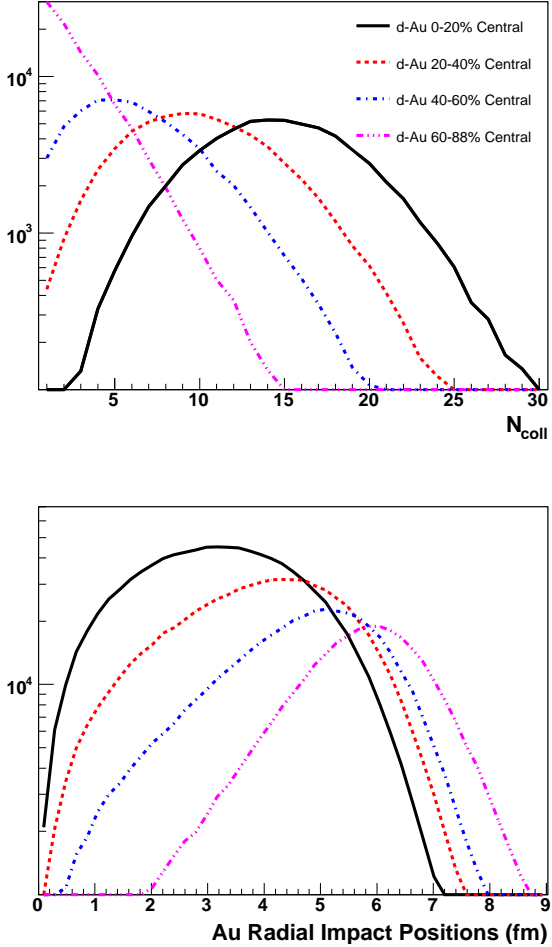}
\caption{\label{fig:nagle_glauber} Results from a Glauber Model Monte
Carlo including simulation of the $\dau$ centrality selection and
triggering based on the PHENIX BBC.  The top panel shows the
distribution of the number of binary collisions for events in each of
the four centrality classes 0-20\%, 20-40\%, 40-60\%, 60-88\%.  The
distribution for radial impact points in the gold nucleus of binary
collisions is shown in the lower panel.}
\end{figure}

A Glauber Monte Carlo and a simulation of the BBC detector used for
centrality determination and triggering are done.  The resulting four
centrality categories (0-20\%, 20-40\%, 40-60\%, 60-88\%) in $\dau$
collisions are characterized by a distribution in the number of binary
collisions, as shown in the top panel of
Figure~\ref{fig:nagle_glauber}.  In addition, the distribution of
radial positions $r$ in the $\rm{Au}$ nucleus of binary collisions is
calculated and shown in the bottom panel of
Figure~\ref{fig:nagle_glauber}.

The procedure is to use the forward, mid, and backward rapidity
centrality-dependent $\rda$ to constrain the modification factor
$\Re(r)$ for three broad regions of initial parton $x$ ($\Re_{low}$,
$\Re_{mid}$, $\Re_{high}$ respectively).  Then one can use these
parameterizations to project the cold nuclear matter effect in the
$\auau$ case.  The current $\dau$ data are insufficient to constrain
the functional form of $\Re(r)$.  As a simplifying case, $\Re(r)$ is
assumed to be linear in $r$ and to be fixed at $\Re(r \ge 8~\rm{fm}) =
1.0$ at the edge of the gold nucleus.  Thus, the only free parameter
is the slope (or equivalently the magnitude of the modification factor
at $r=0$).  Other functions were tried and essentially differ by their
extrapolation to lower and higher radial positions, since the data are
not precise enough to constrain the shape.  This has a particularly
strong impact on the most peripheral collisions for which our
assumption that $\Re(r \ge 8~\rm{fm}) = 1.0$ adds a significant
constraint to the shape.

For all possible slope parameters, consistency with the experimental
data is checked using the procedure detailed
in~\cite{phenix_qhat_constraint_ppg079}, which utilizes the full
statistical and systematic uncertainties.  The range of parameters
within one standard deviation of the uncertainties is determined
separately for backward, mid, and forward rapidity.  Using this range
of parameters, the cold nuclear matter suppression expected in $\auau$
collisions is projected as a function of collision centrality and for
mid and forward rapidity.  Note that the forward rapidity $\auau$
$\jpsi$ production is sensitive to the low-$x$ partons in one gold
nucleus and the high-$x$ partons in the other gold nucleus.  Thus, in
the Monte Carlo, for every binary collision at $r_{1}$ and $r_{2}$
(the radii with respect to the center of each nucleus) the expected
modification is $\Re_{low}(r_{1}) \times \Re_{high}(r_{2})$.  The
midrapidity $\auau$ $\jpsi$ production is predominantly sensitive to
the mid-$x$ partons from both gold nuclei and therefore the expected
modification is $\Re_{mid}(r_{1}) \times \Re_{mid}(r_{2})$.  The total
modification expected is calculated by taking the average over all
correlated $r_1$ and $r_2$ positions for binary collisions within
overall $\auau$ collisions in each $\auau$ centrality class.

\begin{figure}[htb]
\includegraphics[width=1.0\linewidth]{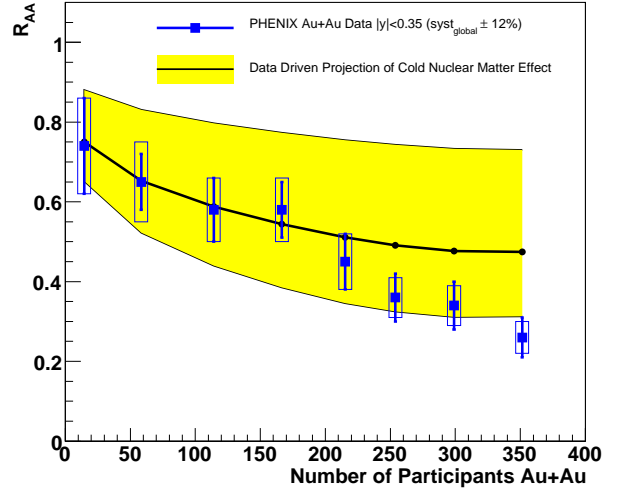}
\includegraphics[width=1.0\linewidth]{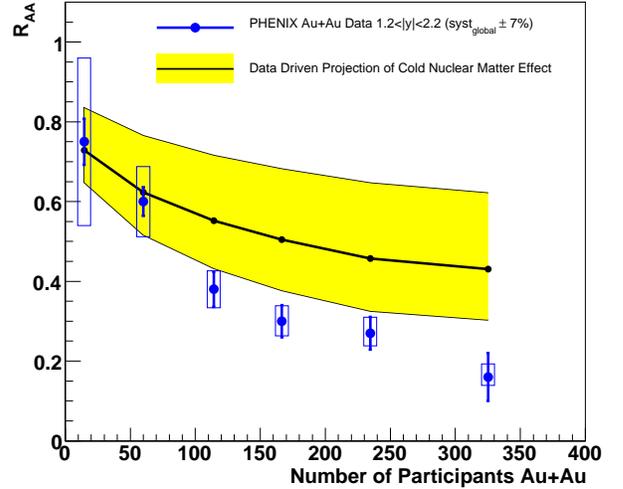}
\caption{\label{fig:nagle} Predictions of the data driven
method~\cite{GranierdeCassagnac:2007aj} constrained by the $\rda$
as a function of collision centrality for the Au+Au $R_{AA}$ for
midrapidity (top) and at forward rapidity (bottom).}
\end{figure}

The results of these calculations matched to the experimentally
measured $\auau$ centrality bins are shown in Figure~\ref{fig:nagle}.
It is notable that the midrapidity cold nuclear matter extrapolation
agrees within the uncertainty of the experimental data at
midrapidity.  Thus, it is not possible within the current constraints
to determine the potential extent of hot nuclear matter effects.  This
conclusion is qualitatively similar to that reached from the previous
model calculations as shown in
Figure~\ref{fig:model_comparisons_auau}.  However, at forward
rapidity, this method projects a somewhat larger range of possible
cold nuclear matter effects than the previous models.

Neither the predictions of cold nuclear matter effects in heavy ion
collisions based on fitting of the $\dau$ data with theoretical curves
(Figures~\ref{fig:model_comparisons_cucu}
and~\ref{fig:model_comparisons_auau}), nor those obtained directly
from the $\dau$ data points (Figure~\ref{fig:nagle}) are well enough
constrained to permit quantitative conclusions about additional hot
nuclear matter effects.

\section{Conclusions}

A new analysis of $\jpsi$ production in \dau~collisions at
$\sqrt{s_{\rm NN}} = 200$ GeV has been presented using the PHENIX 2003
\dau~data set. Cuts and analysis techniques that are consistent with
the previously published results for $\pp$, $\cucu$ and $\auau$
collisions at the same
energy~\cite{Adare:2006kf,Adare:2006ns,Adare:2007cucu} are used.  The
new analysis also benefits from the significantly larger \pp~data set
from Run-5.
%~\cite{Adare:2006kf} when forming the corresponding nuclear
%modification factors.

A statistical comparison of these new results to theoretical
calculations has been performed with a detailed handling of the
experimental uncertainties to estimate a $\jpsi$ (or precursor)
breakup cross section in cold nuclear matter on top of models for the
modifications of the parton distribution functions in the
nucleus. Using EKS (NDSG) shadowing, a breakup cross section of
$2.8^{+1.7}_{-1.4}$ ($2.2^{+1.6}_{-1.5}$) mb is obtained.  These
breakup cross-section values are consistent within large
uncertainties with the $4.2 \pm 0.5$ mb determined at lower energies
at the CERN-SPS~\cite{Alessandro:2006jt}.  The measured values are
then used to predict the expected cold nuclear matter effects on
$\jpsi$ production in $\cucu$ and $\auau$ collisions, and these are
compared to the measured nuclear modification factors for those
systems.  These predictions are found to be similar to those from a
less model-dependent and more data-driven method based on the
variation of the nuclear modification factor measured in $\dau$
collisions as a function of both rapidity and
centrality~\cite{GranierdeCassagnac:2007aj}.  It is notable that the
latter method yields a somewhat larger possible suppression in the
forward rapidity case.  In all cases the large error bars associated
with the extrapolation prevent making firm quantitative statements on
any additional $\jpsi$ suppression in $\auau$ collisions beyond that
expected from cold nuclear matter effects.  A $\dau$ data set with
much improved statistical precision is needed to both reduce the
statistical uncertainties and permit better control over the
systematic uncertainties.

%%%%%%%%%%%%%%%%%%%%%%%%%%%%%%%%%%%%%%%%%%%%%\input{6_Acknowledgments.tex}

%%%%%%%%%%%%%%%%%%%%%%%%%  Acknowledgements 
%\section{Acknowledgements}   % Run-35 long + Ramona for PRC, PLB, etc.

\begin{acknowledgments}
We thank the staff of the Collider-Accelerator and Physics
Departments at Brookhaven National Laboratory and the staff of
the other PHENIX participating institutions for their vital
contributions.  
We also thank Ramona Vogt for useful discussions and for the calculations
used to set the level of the breakup cross sections.
We acknowledge support from the Office of Nuclear Physics in the
Office of Science of the Department of Energy, the
National Science Foundation, Abilene Christian University
Research Council, Research Foundation of SUNY, and Dean of the
College of Arts and Sciences, Vanderbilt University (U.S.A),
Ministry of Education, Culture, Sports, Science, and Technology
and the Japan Society for the Promotion of Science (Japan),
Conselho Nacional de Desenvolvimento Cient\'{\i}fico e
Tecnol{\'o}gico and Funda\c c{\~a}o de Amparo {\`a} Pesquisa do
Estado de S{\~a}o Paulo (Brazil),
Natural Science Foundation of China (People's Republic of China),
Ministry of Education, Youth and Sports (Czech Republic),
Centre National de la Recherche Scientifique, Commissariat
{\`a} l'{\'E}nergie Atomique, and Institut National de Physique
Nucl{\'e}aire et de Physique des Particules (France),
Ministry of Industry, Science and Tekhnologies,
Bundesministerium f\"ur Bildung und Forschung, Deutscher
Akademischer Austausch Dienst, and Alexander von Humboldt Stiftung (Germany),
Hungarian National Science Fund, OTKA (Hungary),
Department of Atomic Energy (India),
Israel Science Foundation (Israel),
Korea Research Foundation and Korea Science and Engineering Foundation (Korea),
Ministry of Education and Science, Rassia Academy of Sciences,
Federal Agency of Atomic Energy (Russia),
VR and the Wallenberg Foundation (Sweden),
the U.S. Civilian Research and Development Foundation for the
Independent States of the Former Soviet Union, the US-Hungarian
NSF-OTKA-MTA, and the US-Israel Binational Science Foundation.
\end{acknowledgments}

%\bibliography{ppg078}

\end{document}